\documentclass[preprint]{aastex}

\newcommand{\p}{$\pm$}

\newcommand{\beam}{beam$^{-1}$}

\newcommand{\msun}{M$_{\sun}$}

\newcommand{\kms}{km s$^{-1}$}

\def\as{$\arcsec$}
\def\am{$\arcmin$}

\slugcomment{Accepted for publication in the Astrophysical Journal Supplement Series, Dec 2010}
\begin{document}
\title{A High-Resolution Survey of HI Absorption toward the Central 200 pc of the Galactic Center}

\author{Cornelia C. Lang}
\affil{Department of Physics \& Astronomy, 703 Van Allen Hall, University of Iowa, Iowa City, IA 52242}
\email{cornelia-lang@uiowa.edu}

\author{W. M. Goss}
\affil{National Radio Astronomy Observatory, Box 0, Socorro, NM 87801}

\author{Claudia Cyganowski}
\affil{Department of Astronomy, University of Wisconsin-Madison, 3321 Sterling Hall, 475 N. Charter Street
Madison WI 53706-1582}

\and
\author{Kelsey I. Clubb\altaffilmark{1}}
\affil{Department of Physics \& Astronomy, Van Allen Hall, University of Iowa, Iowa City, IA 52242}
\altaffiltext{1}{Now at: Department of Physics and Astronomy, San Francisco State University
1600 Holloway Avenue, San Francisco, CA 94132}

\begin{abstract}
We present an HI absorption survey of the central 250 pc of the
Galaxy. Very Large Array (VLA) observations were made at 21 cm in the DnC and
CnB configurations and have a resolution of $\sim$15''(0.6 pc at the
Galactic Center (GC) distance) and a velocity resolution of $\sim$2.5
km s$^{-1}$. This study provides HI data with high spatial resolution, comparable
with the many high resolution observations which have been made of GC
sources over the past ten years. Here we present an overview of the HI
absorption toward $\sim$40 well-known continuum sources and a detailed
comparison of the ionized, atomic and molecular components of the interstellar
medium for the Sgr B, Radio Arc and Sgr C regions. In these well-known  
regions, the atomic gas appears to be closely correlated in both velocity and distribution 
to the ionized and molecular gas, indicating that it resides in photo-dissociation regions
related to the HII regions in the GC. Toward the majority of
the radio continuum sources, HI absorption by the 3-kpc arm is detected, 
constraining these sources to lie beyond a 5 kpc distance in the Galaxy. 
 
\end{abstract}

\section{Introduction}

The bright and unusual radio continuum sources in the central few hundred pc of
the Galaxy provide an opportunity to observe the 21 cm line of atomic
hydrogen in absorption. Previous HI absorption studies toward the
Galactic center (GC) have been crucial in the 
understanding of Galactic structure and rotation and the nature of atomic gas in the inner parts of the Galaxy
(e.g., Schwarz, Shaver \& Ekers 1977; Radhakrishnan \& Sarma 1980; Schwarz, Ekers \& Goss 1982; Liszt et al. 1983, 1985). There are several well-known HI components in the direction of the GC (Cohen \& Davies 1979): 
(1) the expanding 3 kpc arm appears at v=$-$53 km s$^{-1}$ and is thought to be $\sim$5.5 kpc from the Sun, 
(2) HI components near v=+135 \kms~are thought to be beyond the GC 
by distances of a few hundred parsecs to 2 kpc, and (3) the ``nuclear disk'' and ``molecular ring'' components 
appearing at velocities of $-$160 to $-$200 km s$^{-1}$ and $-$135 km s$^{-1}$ are located within a 
few hundred parsecs of the GC. 

In the inner Galaxy, atomic gas is most often associated with regions of molecular gas where 
it serves to shield the molecular gas against photodissociation (Dickey \& Lockman 1990). Therefore, 
HI absorption features not described above may be identified with known GC molecular emission 
features using correlations in velocity structure. The CO survey carried out by Oka et al. (1998) using 
the Nobeyama 45-m telescope 
provides the most favorable resolution, velocity, and spatial coverage of any survey of molecular 
gas within the central Galaxy. In addition, the 
multitude of ``forbidden'' (e.g. sign opposite to galactic rotation) velocity components in the GC region 
are thought to represent the response of the molecular gas in the GC to the Galaxy's strong
stellar bar (Binney et al. 1991; Bally et al. 1988). 

The CO survey data of Oka et al. (1998) illustrated that the molecular gas traced by CO emission 
in the central 200 pc is organized into filamentary and shell-like features. This morphology and kinetic
structure indicates that violent kinetic activity (such as supernova explosions and stellar 
winds from Wolf-Rayet type stars) plays an important role in shaping the ISM. 
In addition to the Radio Arc region (where the Quintuplet and Arches clusters
are located), the GC region is filled with sites where compact thermal radio and mid-infrared
sources have been observed (e.g. Sgr B, Sgr C, and at many positions along the Galactic plane; e.g., 
LaRosa et al. 2000; Egan et al. 1998) and it is likely that massive
stars are either forming or have formed in these regions. In addition, the spectrum of diffuse 
X-ray emission in this region suggests that the ISM is being strongly
influenced by massive star-forming activities (Wang, Gotthelf \& Lang
2002; Wang, Dong \& Lang 2006). 

HI absorption toward the bright SgrA complex and the compact SgrA$^*$ radio
source has been the subject of a number of interferometric studies
over the last three decades (Radhakrishnan et al. 1972, Schwarz, Ekers
\& Goss 1982, Liszt et al. 1983; Dwarakanath et al. 2004). 
These studies illustrate the wide variety of absorption
and emission features toward this complex area of the Galaxy, many of
which have velocities that indicate non-circular motions. On larger scales, 
Lasenby, Lasenby \& Yusef-Zadeh (1989) carried out the first VLA compact
configuration HI absorption study toward the central 30\arcmin~of the GC (corresponding 
to 75 pc at a distance of 8 kpc (Reid et al. 1993)), centered 
on the well-known Radio Arc nonthermal filaments and including the bright SgrA complex. 
The spatial resolution was $\sim$50-70\arcsec~with 
a velocity resolution of 10.2 \kms, with a total velocity
coverage of 660 \kms. However, this spatial resolution of the 
Lasenby et al. (1989) data is 
not adequate for detailed comparisons with higher frequency continuum and
recombination line datasets of the HII regions in the Radio Arc (e.g.,
Lang et al. 1997, 2001) and velocity information on the ionizing stellar
clusters. In addition, the spatial coverage did not include
the active Sgr B region. Although high spatial resolution 
VLA HI absorption measurements have been made toward several 
individual GC sources (e.g., Uchida et al. (1992) and Roy et al. (2003)), a 
complete HI absorption study is missing from the growing canon of
GC surveys.  

Therefore, we have carried out an HI absorption survey of the central 
100\arcmin~$\times$50\arcmin~(250 x 125 pc) of the GC 
in order (1) to understand the physical {\it arrangement} and {\it interactions} between the stellar
and interstellar components, (2) to place constraints on distances to radio continuum features in the field
of view, and (3) to make detailed images and estimates of the HI opacity toward well-studied GC sources.
This current HI absorption survey represents the highest resolution and most complete study of HI absorption toward the GC, and forms the basis for understanding the physical line-of-sight locations and interactions between 
interstellar features in the GC. The 15\arcsec~spatial resolution and a velocity resolution 
2.5 km s$^{-1}$ are a vast improvement over the previous HI absorption studies. In $\S$2 we present a summary of the observations and imaging techniques used to produce the images, in $\S$3, we present the results
and discussion of this HI absorption study. 

\section{Observations \& Imaging}

We have observed at 1.4 GHz (21 cm) using the Very Large Array (VLA) of the National Radio
Astronomy Observatory\footnotemark\footnotetext{The National Radio
  Astronomy Observatory is a facility of the National Science
  Foundation operated under cooperative agreement by Associated
  Universities, Inc.} in the DnC and CnB array configurations.  
The observational details are summarized in Tables 1 and 2. 
In order to cover the $\sim$100\arcmin$\times$50\arcmin~area,  
five fields were observed. The total bandwidth is 1.5 MHz and there
are 127 channels used, corresponding to
$\sim$300 \kms~in velocity coverage. The observations are centered at 0 \kms, and have a
velocity resolution of 2.5 \kms. 
The data were calibrated using the standard {\sf AIPS} routines. 

The line-free channels (channels 10-26) were used to fit a constant
continuum level, and continuum subtraction was performed using the
AIPS task UVLSF. The data were deconvolved jointly using the {\sf
  miriad} software package to take full advantage of the additional
sensitivity from overlapping fields. The continuum mosaic was imaged
using {\it mosmem}, and each plane of the HI absorption
cube was cleaned to a uniform rms level using {\it mossdi} in {\sf
  miriad}. The final spatial resolution of both the continuum and HI absorption line 
images is 15\arcsec. The integration time was 8
hours for the set of five fields in each configuration (i.e., approximately
2.5 hours on each field) which resulted in an rms level of 3-4 mJy beam$^{-1}$ in the
continuum image. The zero level in the continuum image is within 1$\sigma$ (10 mJy) 
of zero, and the image has 
also been corrected for the variable system temperature with frequency in the presence 
of HI emission. The rms noise level in the line images varies between 5-15 mJy beam$^{-1}$. 

\section{Results and Discussion}

\subsection{1.4 GHz Continuum}

Figure 1 shows the continuum image constructed from $\sim$16 line-free channels
at 1.4 GHz. This image is one of the highest-resolution images to date of the more extended
structures in the complex GC region, with excellent sensitivity to both point-like and
diffuse features. Figure 2 is a contour version of the continuum image in Figure 1, with the major
GC sources labeled. In addition, twenty-eight sources 
have been detected in this image and are listed in Table 3. These sources
are categorized as 'compact' or 'extended' and for each source, the position, peak intensity, 
integrated flux density (if resolved), major axes (if resolved), and 
geometric size (square root
of major $\times$ minor axis) are given. These
parameters were obtained by fitting the sources with Gaussian profiles using JMFIT in AIPS. 
Of the 28 continuum sources detected, 16 are classifed as 'compact' and 12 as 'extended'. 
Four of the sources (G0.60-0.20, G0.48+0.07, G0.32-0.19 and
G359.87+0.18) are unresolved and likely to be point sources. We also searched the
compact source catalogs of Yusef-Zadeh et al. (2004) at 1.4 GHz and 
Becker et al. (1994) at 4.9 GHz in order to identify counterparts. Sixteen sources in our
image that have counterparts in one of these two catalogs are listed in Table 4 with
their flux density from our image (1.4-Lang), Yusef-Zadeh et al. (2004; 1.4-FYZ) and
Becker et al. (1994; 5-Becker). 

\subsection{HI Absorption Overview Profiles}

Continuum-weighted, line-to-continuum HI absorption integrated spectra were
produced towards $\sim$40 of the brightest continuum sources in the field. 
The profiles were made to characterize the HI absorption features
toward these well-known sources. In very crowded regions, (e.g., the 
HII regions in the Radio Arc region and SgrA East and West) the AIPS 
task BLANK was used interactively to select the region of interest using
a signal-to-noise cut off in the continuum image. The GIPSY task {\it profil} was used to produce
integrated, continuum-weighted, line-to-continuum spectra towards each
of the 40 continuum features (van der
Hulst et al. 1992). A continuum cutoff of 5$\sigma_{continuum}$ was
used in {\it profil} to obtain optimal signal-to-noise in the
integrated spectra. Since the rms noise varies across the mosaic, a 
value for $\sigma_{continuum}$ was determined locally for each
object. Typical rms noise for the spectra are in the range of 0.01-0.05, where the 
units are line-to-continuum ratio. These profiles were then converted
to profiles of optical depth using the formula $\tau$=-ln(1+(L/C)), where
L/C represents the line-to-continuum profile integrated over a region.  
Multiple Gaussian fits were applied to
each of the profiles (using the GIPSY task {\it xgauprof}), 
and Table 5 lists the parameters of the resulting fits and errors when they can be reliably determined 
(central velocity (v$_{LSR}$), full-width half maximum line width ($\Delta$V$_{FWHM}$), 
and peak opacity $\tau$$_{HI}$). A number of the HI components toward the Sgr B and Sgr A complexes
are highly saturated. Saturated channels will have an undefined
optical depth and occur at the continuum peaks of SgrB and SgrA (i.e., 1-4 Jy/beam respectively)
and in other regions where the line-to-continuum levels are particularly high.

Figures 4-9 show a representative sample 
of 10 HI absorption profiles toward some of the well-known sources in
the GC region: Sgr B1 and B2, Sgr A East and West, the Arched Filaments,
Sickle and Radio Arc, the Sgr C nonthermal filament (NTF) and HII region, 
and two point-like sources in the field: G359.87+0.18 and G359.28-0.26. 
In all figures, there is deep HI absorption near 0 \kms~from many components of 
atomic gas along the line of sight to the GC. 
Components of large HI absorption (line-to-continuum ratios equal to or greater than 0.4)
are apparent in all but the two point-like sources (e.g, Figure 9)
around velocities of $-$50 \kms~and $-$20 \kms. These components are likely to 
be associated with the non-circular velocities of gas near the GC region.  
In addition, most of the profiles show absorption at positive velocities. 
In Sgr B2 and B1, the strong absorption of the dense molecular cloud at 50-60 \kms~is apparent. Toward the Radio 
Arc region (i.e., Sickle, Arched Filaments, Radio Arc nonthermal filaments) there
is little absorption at positive velocities, and most of the absorption 
appears at negative velocities. Both profiles for Sgr A West and East show
HI absorption at positive velocities; Sgr A West has a broad wing between 20 and
100 \kms~which is discussed in $\S$3.7, and atomic gas associated with the dense 
molecular streamers at v=20 and 50 \kms~near Sgr A East are apparent in Figure 7 (top)
(e.g, McGary et al. 2001; Coil et al. 1999). Figure 8 shows the HI absorption 
toward parts of the Sgr C complex: atomic gas associated with various molecular features 
near $-$100 and $-$60 \kms~are evident.  Finally, the profiles in Figure 9 show
the HI absorption toward two point sources in the field. 
A more detailed discussion of the HI and its relationship to the other 
components of the interstellar medium in the GC for the main complexes (Sgr B, Radio Arc,
and Sgr C) follows below in Sections 3.4, 3.5 and 3.6. 

\subsection{Large-Scale Negative Velocity HI Absorption}

As mentioned above, prominent HI absorption occurs toward most bright continuum sources in 
this field at a number of negative velocities. Table 6 summarizes the negative velocity
components and lists which continuum sources toward which the negative velocities are detected.
A listing of four different clouds (A-D) is given in this table as well as the average velocity and 
rms noise, the FWHM and rms noise, and the average optical depth and rms noise 
of each component. The atomic gas associated with the 3 kpc-arm is present 
in Component A. Only 8 of the 40 continuum sources do not have absorption in
the range of $-$54 to $-$60 \kms~(i.e., Component A): sources 1, 2, 10, 11, 19, 27, 31, 40. 
In addition, Figures 10 and 11 show the distribution of HI opacity at $\sim$ $-$25 \kms~(Component C) and
$-$54 \kms~(Component A), respectively, across the entire field of our HI study. The greyscale 
shows HI opacity and is overlaid on contours of molecular gas (CO (1-0) from Oka 
et al. (1998)) that define the orientation of the Galactic plane.  
The negative velocity components C and D (at velocities of $-$41 to $-$20 \kms) 
are likely to be located within the central few hundred parsecs of the Galaxy
where the gas motions are ``forbidden'' and may result from an inflow toward
the GC region from the barred-potential distribution (e.g., Binney et al. 1991).
This pattern is in agreement with the conclusion about the negative velocity molecular
clouds that were detected toward the GC in the H$_2$CO study of Mehringer et al.
(1995).  

\subsection{HI Absorption toward the Sgr B Complex}

Sgr B2 is one of the most massive star forming regions in the Galaxy.
In addition to an extensive, dense molecular cloud ($>$ 5 $\times$ 10$^6$ \msun), a
large number of compact and ultra-compact HII regions and masers
indicate that this region is young with many embedded sources. 
Higher resolution radio observations of Sgr B2 North (Sgr B2 (N))and Sgr
B2 Main (Sgr B2 (M)) reveal ~50 compact sources at
sub-arcsecond level (Gaume et al. 1995;
DePree et al. 1995, 1996). On the other hand, the radio emission 
arising from the Sgr B1 complex
indicates that this region is a more highly evolved
star-forming region than Sgr B2 with numerous filamentary and shell-like ionized
structures (Mehringer et al. 1992).

The ionized and molecular gas in both regions have been studied at
high spatial resolutions (e.g., Mehringer et al. 1993, Mehringer et 
al. 1992, Mehringer et al. 1995). The ionized gas in Sgr B2 has 
velocities between 50-70 km s$^{-1}$ and the ionized gas in Sgr B1
has velocities between 30-50 \kms. The Sgr B1 and B2 complexes are 
thought to be physically related and the velocity distribution across both 
of the molecular clouds supports this 
hypothesis. Absorption of H$_2$CO (formaldehyde) gas toward Sgr B1 
reveals that the molecular gas has a more extended and shell-like distribution 
than in Sgr B2 (Mehringer et al. 1995). In Sgr B2, the H$_2$CO opacities are
very high (e.g., $\tau$ $>$ 1 in many places and $>$ 5) and are often 
correlated in velocity with the radio recombination line data of Mehringer 
et al. (1993), suggesting that there may be compact HII regions embedded in 
this cloud (Mehringer et al. 1995). 

G0.6-0.0 lies in projection between the prominent Sgr B2 and Sgr B1 complexes. 
The recombination line study of Mehringer et al. (1992) indicates that both
kinematically and morphologically G0.6-0.0 is physically related to
the Sgr B1 and Sgr B2 complexes. 
Figure 12 shows a 1.4 GHz continuum image of the Sgr B region from this
study. Small regions for which HI absorption spectra were produced are
labeled according to the nomenclature of Mehringer et al. (1993) for 
Sgr B2 and Mehringer et al. (1992) for Sgr B1 and G0.06-0.0. Table 7 lists
fits to various interstellar components (HI, HII and H$_2$CO) for each 
of these regions. 

\subsubsection{Sgr B2}

The HI absorption is very strong across Sgr B2 at velocities
of 50$-$80 \kms. Figure 13 shows the distribution of HI opacity 
for the Sgr B2 N and M components in greyscale ($\tau$$_{HI}$ shown for
values of 0$-$2) and contours representing the continuum emission. This plot was
produced by making a cube of the distribtion of HI opacity 
for each pixel. The HI line absorption is very
strongly saturated (i.e., $\tau$ $\sim$ 3-4) in these regions. 
Figure 14 shows the central velocity distribution of HI integrated
over small regions (defined by the box size in this figure) in the 
Sgr B2 complex. Figures 15-18
show profiles of HI opacity for six regions in the Sgr B2 complex and
columns 2$-$4 of Table 7 list the central line velocity (v$_{HI}$), opacity 
($\tau$$_{HI}$) and $\Delta$V (FWHM line width) fits for these profiles. Columns
5 and 7 give the central velocity of the H$_2$CO absorption and ionized gas 
(HII), respectively, from Mehringer et al. (1995, 1993). 

These HI absorption data reveal some patterns: toward Sgr B2 (N), 
the HI absorption occurs at higher velocities (65$-$70 \kms) than toward
Sgr B2 (M), where the velocities are $\sim$ 60-65 \kms. In addition, 
Sgr B2 (R) and Sgr B2 (S) have HI velocities between 50$-$60 \kms.
The ionized gas from the H110$\alpha$ ratio recombination line study of  
Mehringer et al. (1993) shows that the ionized gas in the different 
components of Sgr B2 is very well-matched to the atomic gas. 
In fact, Table 7 shows that for the sources in Sgr B2, there is
very close agreement between the velocities of HI, H$_2$CO and HII 
gas in regions 18, 20, 15 and 17. 

The current paradigm for the star formation occuring in the Sgr B2 complex 
is that molecular cloud collisions
have likely triggered star formation throughout the region (Mehringer et
al. 1993, 1995; Hasegawa et al. 1994, 2008). The well-known Sgr B2
molecular cloud has a velocity of $\sim$65 \kms; however, the range of
observed velocities is approximately 35-95 \kms~(Hasegawa et al. 1994). 
H$_2$CO absorption has revealed that there is a higher velocity cloud 
(v $\sim$80 \kms) present that is thought to have been the trigger for 
star formation in this region. Interestingly, ionized gas is found
in the vicinity of this molecular cloud at velocities 
that are intermediate between the two clouds (H110$\alpha$ velocities 
$\sim$70 \kms~compared with $\sim$65 \kms~and $\sim$80 \kms~of the molecular gas; 
Mehringer et al. 1993). 
Variations in the opacity and kinematics of H$_2$CO absorption and radio
recombination lines over the Sgr B2 source indicate that sites of
active star formation (HII regions) are embedded within the cloud at
different line-of-sight distances. 

The presence of HI and H$_2$CO absorption at higher velocities toward Sgr B2 (N)
may indicate that this source may be located deeper in the cloud than
regions (such as R, Main) where there is no absorption at higher
velocities. This pattern was first suggested by Mehringer et al. (1995). 
Toward Sgr B2 (R) there is HI absorption at relatively low
velocities (~55 \kms). The absence of HI absorption at higher velocities
from the molecular cloud on the backside of the Sgr B2 molecular cloud
indicates that this source is likely to be on the nearest side of the
complex. The cartoon in Figure 20 shows a sketch of a proposed 
line-of-sight arrangement of the atomic, ionized and molecular
components in Sgr B2.

For the positions where both the HI absorption and H$_2$CO absorption 
lines have good signal-to-noise, it is possible to make an estimate of the
ratio of the number density of HI (N$_{HI}$) to the number density of 
molecular hydrogen (N$_{H_2}$). Mehringer et al. (1995) provide the 
ratio of the column density of H$_2$CO molecules to the excitation 
temperature (N$_{L}$/T$_{ex}$), which depends on the optical depth of 
the H$_2$CO emission and the line width. The excitation temperature
for H$_2$CO is assumed to be 1-2 K (Mehringer et al. 1995). 
The column density of H$_2$ can
then be inferred by the ratio N$_{H_2CO}$/N$_{H_2}$ = 2 $\times$10$^{-9}$. 
For the strong lines in Sgr B2 complex, the values for column density of H$_2$ 
are N$_{H_2}$=7$-$30 $\times$10$^{22}$ cm$^{-2}$. To derive the
ratio of N$_{HI}$ to N$_{H_2}$, we need a measure of N$_{HI}$, obtained by using
the standard formula N$_{HI}$ = 1.92 $\times$ 10$^{18}$ T$_{ex}$ $\tau_{HI}$ $\Delta$V. 
Here, T$_{ex}$ for HI is not well known for the GC environment; we have
assumed T=60 K. Values for N$_{HI}$ in Sgr B2 therefore range from 
2$-$6 $\times$ 10$^{21}$ cm$^{-2}$, producing ratios of 
N$_{HI}$ to N$_{H_2}$ of 0.02-0.05. These values are consistent with similar
values obtained by Lasenby et al. (1989) in their HI absorption study
of the central 50 pc. Table 7 lists the N$_{HI}$ to N$_{H_2}$ ratio in column 6.
In addition, the close correlation in distribution and central velocity of
the HI absorption and the ionized gas also suggests that atomic gas 
resides in photo-dissociation regions related to the HII regions in the GC 
similar to the scenario in both W3 and Orion B (e.g., 
van der Werf \& Goss 1990; van der Werf et al. 1993). 

\subsubsection{Sgr B1 and G0.6-0.0}

Sgr B1 can be divided into an Eastern component (Sgr B1-East) and 
Western component (SgrB1-West). In previous
studies of both ionized and molecular gas, these regions are also known to be kinematically
distinct (Mehringer et al. 
1992, 1995). The integrated HI opacity profiles for these regions 
are shown in Figure 18, and fits to these profiles are listed in Table 7. 
The main positive-velocity HI feature in Sgr B1-East is centered at $\sim$47 \kms, 
with $\tau$$\sim$2. The molecular gas absorption
has a similar velocity ($\sim$+50 \kms; Mehringer et al. 1995). 
Using the H$_2$CO absorption feature near +50 \kms, we calculate 
the ratio of N$_{HI}$ to N$_{H_2}$ (as above) and obtain a value of $\sim$0.1,
about 2-3 times as large as the values in Sgr B2 (0.02-0.05). 
The velocities of the ionized gas in Sgr B1-East are +30 \kms, and at +60 \kms, 
not as well-correlated to the velocities of the atomic and molecular material. 

The HI absorption profile toward Sgr B1-West shows two components 
at +20 \kms~and +40 \kms, with values of $\tau$ of $\sim$0.5. The ionized gas toward this 
region is in the range of +40 to +45 \kms, and the H$_2$CO absorption
occurs over the range of +30 to +40 \kms. The atomic gas is
therefore likely to be associated with the ionized and molecular
gas in the +40 \kms~cloud. Due to the low signal to noise in both the HI and H$_2$CO profiles, 
the N$_{HI}$ to N$_{H_2}$ ratio for this region is quite uncertain. 
Mehringer et al. (1995) present a scenario for the Sgr B1-East
and Sgr B1-West HII regions in which Sgr B1-West HII region is on the nearside of
the Sgr B1 molecular cloud (with v=+40-50 \kms) and the Sgr B1-East HII region lies on the backside of the molecular cloud. 
The lack of H$_2$CO absorption 
near +50 \kms~toward Sgr B1-West indicates to Mehringer et al. (1995) 
that this feature is located on the nearside of the cloud. 
The HI profiles presented here appear to follow this general 
pattern, with no HI absorption at +50 \kms~toward Sgr B1-West, 
and the atomic gas more closely associated in velocity to 
the ionized component. 

Figure 19 shows the HI opacity toward the source G0.6-0.0. 
The prominent HI absorption occurs at $\sim$52 \kms. 
The velocities of the atomic, molecular and ionized gas 
in this region agree well at approximately +51-55 \kms.  
These values lie between the range for Sgr B1 ($\sim$+40 to +50 \kms)
and Sgr B2 ($\sim$+60 to +70 \kms), further indicating that G0.6-0.0 may be
a kinematically-intermediate region between Sgr B1 and Sgr B2. 
The N$_{HI}$ to N$_{H_2}$ ratio
for G0.6-0.0 is 0.2. This order of magnitude difference from the ratio
measured in Sgr B2 may be due to the lower column of molecular gas in this region. 
The HI column densities for both G0.6-0.0 and Sgr B1 regions are
similar (but are typical of the lower end of the range) to those in Sgr B2.   
In general, the HI optical depths are lower toward the Sgr B1
components ($\tau$$\sim$0.5-1.5) than in Sgr B2 ($\tau$ $>$2 or more). These lower
values are also consistent with the idea that the material related to 
star formation (i.e. molecular gas and the associated ionized
layers) is more dispersed in Sgr B1 and G0.6-0.0 and that these
sources are more evolved than Sgr B2. 

\subsection{HI Absorption toward the GC Radio Arc}

The GC ``Radio Arc'' region consists of a 
number of thermal and non-thermal features. Figure 21 shows
the 1.4 GHz continuum image from these data. The major
features of this region are labeled: (1) the linear, non-thermal
filaments (also known as NTFs) which extend for 16\arcmin~(40 pc), first discovered
by Yusef-Zadeh et al. (1984), (2) the thermal Arched Filaments
HII regions which have been studied in detail by Lang et al. (2001, 2002)
and are assumed to be ionized by the massive Arches cluster (Figer et al. 2002), 
(3) the thermal Sickle HII region and Pistol Nebula, known to be associated with the Quintuplet
cluster (Figer et al. 1999), and (4) the ``H''-regions, a set of 5 compact
HII regions that are located in projection between the SgrA
complex and the Arched Filaments (Yusef-Zadeh \& Morris 1987; Zhao et al. 1993).
There are two well-known molecular clouds in the vicinity of the
Radio Arc that are thought to be physically associated with the
radio continuum sources: the ``-30 \kms~cloud'' and the ``+25 \kms~cloud'' (Serabyn \& Guesten 1987; 
Serabyn \& Guesten 1991). Both clouds have forbidden velocities, likely due to their highly
non-circular orbits around the GC. 
Therefore, we have examined the HI absorption in detail near these
velocities.  Figures 22 and 23 show contours of 1.4 GHz continuum emission overlaid
with greyscale representing the HI opacity for 
velocities around $-$30 \kms~(Figure 22) and +25 \kms~(Figure 23)
toward the Radio Arc region. 

\subsubsection{Arched Filaments HII Complex}

Figures 24-26 show continuum-weighted, integrated profiles of 
HI opacity for small regions ($\sim$1-2\arcmin) in the Arched Filaments HII complex. 
These regions are labeled according to their location in the Arched Filaments, 
which is shown in Figure 21: the Eastern portions (Arches-E1, Arches-E2) 
and Western portions (Arches-West, Arches-W1 and Arches-W2). 
Table 8 lists the fits to the HI components that have the highest signal-to-noise, 
as well as the central velocities of the ionized and molecular gas from
previously-published studies (e.g., Serabyn \& Guesten 1987; Lang et al. 2001). 

The velocities of the ionized, atomic and molecular gas 
components are in a narrow range, primarily 
between $-$28 \kms~and $-$55 \kms, with much of all three
components near $-$30 to $-$40 \kms. To further illustrate 
the velocity structure of the 
HI absorption, Figure 27 shows the central velocity 
for the HI determined from opacity profiles made over
even smaller regions (indicated by the squares in Figure 27). 
Here, it is also clear that much of the HI component lies
at velocities of $-$25 to $-$35 \kms. 
Figure 28 shows a cartoon from the paper by
Lang et al. (2002), where the authors suggest an arrangement of
the ionized, molecular and atomic gas and how this interstellar
material may be arranged around the ionizing cluster. 
The velocities of the HI absorption from this study and
as compared with the ionized and molecular gas from previous
work appear to be consistent with this scenario. 

\subsubsection{Sickle HII Region, Pistol Nebula and Radio Arc Nonthermal Filaments}

Figures 29-31 show continuum-weighted, integrated profiles of 
HI opacity for small regions ($\sim$30\arcsec~$\times$30\arcsec) in 
the Sickle HII region and Pistol
Nebula. Table 8 lists the fits to the HI components where
the signal-to-noise is the highest. In addition, central
velocities are listed for the ionized and molecular gas in this
vicinity (due to the ``+25 \kms~cloud'') from previously-
published studies (e.g., Serabyn \& Guesten 1991; Lang
et al. 1997). Unlike in the Arched Filaments complex, the 
correlation between atomic, ionized and molecular gas is
not as strong. 
In fact, both the ionized and molecular gas are dominated by positive 
velocities (e.g., near +25 \kms) while the atomic gas is concentrated
at negative velocities (e.g.,  $\sim$ $-$30-$-$50 \kms). One 
possible explanation for the $\sim$50 \kms~difference in the velocities
of the interstellar components is if the atomic gas lies
in a shell-like distribution in front of the Radio Arc and
is expanding away from this region. Such an arrangement could be related
to the episodes of massive star formation that are ongoing in the
GC Radio Arc (Wang, Dong \& Lang 2006) and is in agreement with a
similar suggestion by Lasenby et al. (1989). In general, these
observations compare very favorably to the earlier HI absorption
study of Lasenby et al. (1989), which covered the Radio Arc
region in detail. Here, the velocity resolution of the HI data is
improved, and the corresponding ionized and molecular line observations
have higher spatial resolutions. However, the conclusions are
similar: much of the atomic gas in this region appears to lie
on the near side of the continuum arc and filaments in the 
Radio Arc. Figure 31 shows a continuum-weighted, integrated profile of 
HI opacity for a region of the NTFs in the Radio Arc that lie
east of the Sickle and Pistol. The main HI absorption occurs
near $\sim$-50 \kms~and $-$30 \kms, indicating that there is 
atomic material on the near side of the NTFs at these ``forbidden''
velocities. This atomic material is likely to be physically
related to several clouds in the region at similar velocities
(i.e, the $-$30 \kms~cloud). 

\subsubsection{'H'- regions: HII regions}

Figures 32 and 33 show continuum-weighted, integrated profiles of 
the HI opacicty toward the compact HII regions known as 
H1, H2, H3 and H5. Table 8 lists the fits to the HI components 
where the signal-to-noise is the highest. The table also lists
the central velocities of the ionized and molecular gas from
previously-published studies (i.e., Zhao et al. 1993; Serabyn
\& Guesten 1987). Especially noticable is the close
correlation between the ionized, molecular and atomic components
in these compact HII regions. 
The similarity in velocities of the ionized and molecular gas in
the HII regions suggest that they are part of the larger molecular
gas reservoir that is associated with the Arched Filaments complex.

\subsection{HI Absorption toward the Sgr C Complex}

The Sgr C complex (G359.45-0.05) is located $\sim$30\arcmin~(or 75 pc) 
in projection to the South of the Sgr A complex, along the Galactic plane.
Figure 34 shows the 1.4 GHz continuum image of the main components of
the Sgr C complex: the SgrC HII region, the NTF, 
and HII regions known as FIR 4, Source C and D (\citet{LS95}, \citet{R03}, and 
\citet{OF84}). The NTF has been labeled ``Part A'' and ``Part B'' after Roy 
(2003). 
There are similiarities between Sgr C and the Radio Arc: the Sgr C HII region
may contain more than 250 solar masses of ionized gas with 
massive stars present, including some late-type O stars.  This
HII region is associated with a molecular cloud which has the same
velocity ($\sim$$-$65 \kms).  The molecular cloud
associated with Sgr C appears to be organized in a shell-like distribution 
surrounding the HII region; the massive stars are thought to have blown  
a cavity in the distribution of gas (Lizst \& Spiker 1995). Finally, the 
Sgr C NTF is also thought to be physically related to this star formation
region. 

Figures 35-37 show the corresponding integrated HI absorption profiles
for each of these regions: Part A and B of
the Sgr C NTF, the SgrC HII region, FIR4, and Sources C and D. 
Most of the profiles have a
complicated velocity structure with absorption at many velocities.  
As mentioned above, there are molecular clouds in this region with
velocities of $-$65 \kms~and $-$100 \kms, noted by \citet{LS95}
and \citet{O01}. Figure 38 shows contours of 1.4 GHz radio continuum emission
overlaid on greyscale representing the HI optical depth at velocities 
of $-$65~\kms~and $-$100~\kms. 
Figure 39 shows the same HI optical depth as in Figure
38 overlaid on contours of CO (J=1-0) emission from the survey of Oka et al. (1998) at 
velocities of $-$65 \kms~and $-$100 \kms. 

\subsubsection{Sgr C NTF}

Figure 35 shows the HI absorption profile toward Part A and Part B 
of the Sgr C NTF. These profiles are similar with 
HI absorption near $-$60-65 \kms~toward both regions of
the NTF, but clearly stronger toward Part B of the NTF.
This difference in the HI absorption by the $-$65 \kms~cloud across the 
Sgr C NTF is striking and may indicate that the atomic gas in 
the direction of Part A is embedded in or located on the backside of the 
$-$65~\kms~molecular cloud. 
In addition, there is HI absorption near $-$100 \kms~in both profiles. 
In Part B, the HI absorption occurs at $-$136 and $-$109 \kms; 
both components may be associated with the $-$100 \kms~molecular cloud 
if there is a velocity gradient. 

Figures 35 and 38 (right) illustrate that the HI absorption by the $-$100 \kms~cloud 
is stronger toward Part A of the NTF ($\tau$$_{HI}$=0.15) than toward 
Part B of the NTF ($\tau$$_{HI}$=0.45). 
In addition, Figure 39 (right) shows that the distribution of molecular gas traced in the
CO emission near $-$100 \kms~peaks near Part A of the NTF and is nearly
absent near Part B of the Sgr C NTF. This correlation of the stronger
HI absorption at $-$100 \kms~with the peak of molecular gas suggests that 
the atomic gas is physically associated with the molecular cloud at this
location and lies behind the NTF. 
The change in HI absorption across the NTF (from $\tau$$_{HI}$=0.45 to 0.15 
from Part A to Part B) indicates that the atomic and molecular material
may be located partially in front of the NTF 
as the continuum brightness of the NTF does not vary substantially across
these regions. The NTF may be embedded in or located 
on the backside of the $-$100~\kms~molecular cloud. 
Figure 40 illustrates the a possible arrangement of interstellar sources 
in this region (and in particular, the arrangement of the molecular
cloud and NTF). 

\subsubsection{Sgr C HII Region}

Figure 36 (left) shows that the HI absorption toward the Sgr C HII region is strongest near $-$60 \kms,
also the velocity range of the known CO cloud associated with the Sgr C 
HII region (Liszt \& Spiker 1995). The radio recombination line spectrum of the HII region 
is centered on $-$65.5 \kms. Therefore, it is likely that the HI at this velocity is
physically associated with this complex. Further, we can obtain some
details about location relative to the molecular and ionized components
by comparing the distribution of each component. 
Figure 39 (left) shows that the Sgr C HII Region 
appears to be located in a cavity of the $-$65~\kms~molecular material; there
is a striking, exact anti-correlation between the atomic/HII and molecular 
components. The absorption against the Sgr C HII region indicates that there
is atomic material on the near side of the HII region, probably mixed in 
with the ionized gas. The molecular material may presumably have been ionized
and blown out of this cavity. It is less clear that there is $-$100 \kms~HI absorption toward the Sgr C 
HII region; in fact, the absorption near $-$100 \kms~shifts to $\sim$120 \kms~near
the Sgr C HII region and may not be part of the same cloud 
(in agreement with \citet{LS95}) and the $-$100~\kms molecular 
material is only observed towards Part A of the Sgr C NTF in Figure 37 (right). 

The Sgr C NTF and the Sgr C HII Region both appear to be associated with the $-$65~\kms molecular 
cloud in the Sgr C region, which suggests that the NTF and HII region themselves are physically related. 
This possible association between the HII region and NTF is particularly relevant
to the mechanism for accelerating particles along the NTFs, an open issue in
understanding the nature of these unusual non-thermal filaments. 
\citet{SM94} have previously suggested that the relativistic particles 
present along the NTFs originate in the ionized edges of molecular clouds, based
on their finding of molecular clumps coinciding with NTFs in the Radio Arc. Therefore,
the scenario in Sgr C may be consistent with this picture. 

\subsubsection{HII Regions Surrounding Sgr C and Line-of-Sight Arrangement}

Figure 36 (right) and Figure 37 show the HI absorption toward
three HII regions adjacent to Sgr C: FIR 4, Source C and D. 
Due to the low surface brightness of FIR 4 in the 1.4 GHz continuum, the HI absorption data
has low signal to noise (Figure 36 (right)). The prominent features in Figure 36 (right) are HI 
absorption near $-$80 \kms~,$-$54\kms, and $-$135 \kms. The lack of absorption 
due to either the $-$100~\kms~molecular cloud or the
$-$65~\kms~molecular could indicate that FIR 4 could be located in 
front of these clouds but beyond the 3-kpc arm at 5 kpc, since the $-$54 \kms~absorption
component is present. Source C has an absorption spectrum which appears very similar to the
other sources in the Sgr C complex, with absorption near $-$60~\kms, $-$100 \kms~
and $-$126 \kms. By contrast, Source D only has HI absorption features in the range
of $-$30 to 30 \kms, suggesting that it may be nearer to the Sun along the line
of sight. 

Using the comparison of atomic absorption and molecular and ionized gas velocities, we can
make a schematic of the line-of-sight arrangement of the sources and ISM surrounding Sgr C. A possible 
arrangement from the results presented in this paper is shown in Figure 40
(as viewed from above, looking down on the region, with the observer to the left in the drawing). 
We make the following assumptions: Source D is a foreground source located somewhere between the 
Sun and the 3~kpc~arm, not associated with the rest of the Sgr C region. The atomic gas associated
with the $-$65 \kms~molecular cloud is likely to be located in front of the Sgr C HII Region, Source C and 
likely to be physically associated with the Sgr C NTF. The atomic gas associated with the 
$-$100 \kms~molecular cloud is also likely to be located on the near side of the Sgr C NTF (and 
possibly physically related) and of Source C. The results presented here (including the line-of-sight
arrangement of components associated with Sgr C) agree very well with those derived in Roy (2003) using the GMRT. 
The spatial and velocity resolutions are very comparable; our survey differs in that it covers
a large region whereas the Roy (2003) observations were targeted at several NTFs.

\subsection{HI Absorption toward Sgr A: High Spatial Resolution Data}

Complementary HI observations of the Sgr A field were carried out by Dwarakanath, Goss, Zhao and Lang (2004) 
using the VLA in the A, B, C and D arrays in January and October of 2002. The purpose of 
these observations was to elucidate the nature 
of the wide line ($\Delta$V$_{FWHM}$ $\sim$ about 100 \kms) toward Sgr A. These complementary observations
have a much larger velocity coverage of $\sim$600 \kms~and a velocity resolution of 1.3 \kms. 
The wide velocity range was required to cover the large number of velocity components at the GC, 
while the high velocity resolution was required to identify and remove the narrow HI absorption lines. 
Figure 41 shows the HI spectra obtained with an angular resolution of $\sim$8\arcsec~at four positions 
in SgrA West: The spectrum 'a' is toward SgrA* while 'b' is displaced by 35\arcsec~to the SW. 
Positions 'c' and 'd' are located 2\arcmin~E and NE of SgrA* and within the SNR SgrA East. 
A number of narrow lines ($\sim$10 \kms) are obvious. In the 'b' direction a broad shoulder is observed, implying a wide component with a $\Delta$V$_{FWHM}$ of $\sim$120 \kms. At positions 'c' and 'd', the wide line is not detected. The wide line is interpreted 
to originate in various HI components of the circumnuclear disk ('CND') with a radius of 1.3\arcmin~centered on SgrA*. These components have central velocities of $\sim$100 \kms~to the N of Sgr A* and $\sim$$-$100 \kms to the S. The observed wide line in the direction of SgrA does not represent a component of low opacity, turblent shocked HI clouds; this feature is the result of the superposition of various components of the CND surrounding SgrA* (e.g., see Dwarakanath et al. 2004 for more details).

\subsection{Constraining Line-of-Sight Distances to GC Sources}
 
The majority of continuum sources in this study have absorption due to the ``3-kpc arm'' (at a velocity of $\sim$$-$54 \kms~and at a distance of $\sim$5 kpc). Table 5 lists the presence of this absorption feature in each spectrum. The presence of this feature suggests that these sources must lie beyond 5 kpc. Additionally, many of these sources are associated with forbidden-velocity molecular clouds that are believed to lie at the GC. The HI spectra, therefore, of the majority of the sources in Table 5 are consistent with the sources being at a GC distance of d=8 kpc. Only three sources (G0.32-0.19, G0.31-0.20, and G359.28-0.26) do not show absorption by the 3-kpc arm. The absence of this feature suggests that these sources may lie between the Sun and the 3-kpc arm and we consider these to be foreground sources. Finally, the extragalactic source G359.87+0.18 has been studied by Lazio et al. (1999) and our HI spectrum for this source is in agreement with these findings. 

A website has been created to disseminate the images, profiles and data to the 
astronomical community (see http://astro.physics.uiowa.edu/$\sim$clang/gchi). 

\begin{acknowledgements}
The authors would like to thank K. S. Dwarakanath for his assistance with putting together 
the catalog and analyzing the HI profiles, and Sungeun Kim for assistance with the initial imaging. The authors
also thank Crystal Brogan for her advice and assistance.  
C.C. L. would also like to acknowledge the dedication and hard work of her students Chrissy Roark, Sarah Willis, and
Emily Richards for assembling the initial continuum and line catalog and fitting HI absorption profiles.   
\end{acknowledgements}

\begin{deluxetable}{lccccc}
\tablecaption{Summary of Observations}
\tablewidth{0pt}
\tablehead{
\colhead{Field}& 
\colhead{Date}&
\colhead{Array}&
\multicolumn{2}{c}{Phase Center}&
\colhead{Integration Time}\\
\cline{4-5}
\colhead{}&
\colhead{}&
\colhead{}&
\colhead{$\alpha$ (J2000)}&
\colhead{$\delta$ (J2000)}&
\colhead{(hours)}}
\tablecolumns{5}
\startdata
GCHI-1 	&Sept 2001	&DnC	&17 47 00.3	&$-$28 32 06	&1.6\\
$...$	&Jun 2001	&CnB	&$...$	&$...$	&4\\
GCHI-2 &Sept 2001&DnC&17 46 24.5&$-$28 44 55&1.6\\
$...$	&Jun 2001	&CnB	&$...$	&$...$	&4\\
GCHI-3 &Sept 2001&DnC&17 45 48.9&$-$28 57 44&1.6\\
$...$	&Jun 2001	&CnB	&$...$	&$...$	&4\\
GCHI-4 &Sept 2001&DnC&17 45 13.1&$-$29 10 32&1.6\\
$...$	&Jun 2001	&CnB	&$...$	&$...$	&4\\
GCHI-5 &Sept 2001&DnC&17 44 37.2&$-$29 23 20&1.6\\
$...$	&Jun 2001	&CnB	&$...$	&$...$	&4\\
\enddata
\end{deluxetable}

\begin{deluxetable}{lc}
\tablewidth{0pt}
\tablecaption{Parameters of the HI Line Observations}
\tablehead{\colhead{Parameter}&\colhead{HI~Data}}
\tablecolumns{2}
\startdata
HI~Rest Frequency&1420.406 MHz\\
LSR Central Velocity     &0 \kms \\
Total Bandwidth  &1.5625 MHz (330 \kms)\\ Number of Channels      &127\\
Channel Separation        &12.2 kHz (2.5 \kms)\\
Velocity Coverage&330 \kms\\
Flux Density Calibrator &3C286\\
Bandpass Calibrator& NRAO 530 (1730-130)\\
Phase Calibrator        &1751-253\\
\enddata
\end{deluxetable}

\begin{deluxetable}{lcccccc}
\tabletypesize{\footnotesize}
\tablecaption{1.4 GHz Continuum Compact and Slightly Extended Sources}
\tablehead{\colhead{Source} &
\colhead{Source} &
\colhead{RA} &
\colhead{DEC} &
\colhead{$I_p$} &
\colhead{$S_{1.4}$} &
\colhead{Geom. Size} \\
\colhead{Name} &
\colhead{Type} &
\colhead{(h m s)} &
\colhead{(d \am \as)} &
\colhead{(mJy \beam)} &
\colhead{(mJy)} &
\colhead{(arcsec)}}
\tablecolumns{7}
\startdata
G0.60-0.20     &Compact       &17 47 50.23   &-28 31 25.3   &31\p10\tablenotemark{\dag}         &$...$ & $...$\\
G0.59+0.05     &Extended      &17 46 49.90   &-28 24 10.9   &55\p9             &454\p84     &40 \\
G0.59-0.13     &Compact       &17 47 30.17   &-28 29 57.4   &43\p10     &87\p28  &15\\
G0.53+0.18     &Compact       &17 46 10.01   &-28 23 24.7   &315\p10        &504\p24     &11\\
G0.53+0.13     &Extended      &17 46 22.06   &-28 25 03.0   &36\p10     &106\p36     &20     \\
G0.50+0.17     &Compact       &17 46 08.30   &-28 25 31.7   &46\p10     &133\p36     &21  \\
G0.49+0.19     &Extended      &17 46 02.39   &-28 24 50.0   &\phn68\p9     &368\p59     &31\\
G0.48+0.07     &Extended      &17 46 28.21   &-28 29 15.6   &80\p10\tablenotemark{\dag}         &$...$& $...$  \\
G0.48-0.10     &Compact       &17 47 08.20   &-28 34 48.3   &46\p10        &64\p22  &10  \\
G0.47-0.10     &Compact       &17 47 06.00   &-28 35 11.2   &44\p10     &82\p26  &13    \\
G0.43-0.06     &Compact       &17 46 51.39   &-28 36 09.1   &53\p10        &72\p21  &9  \\
G0.38+0.02     &Extended      &17 46 27.79   &-28 36 03.6   &422\p9        &1545\p43        &24    \\
G0.35-0.03     &Compact       &17 46 32.68   &-28 39 14.1   &89\p10     &178.3\p28     &15     \\
G0.33-0.01     &Extended      &17 46 27.65   &-28 39 25.0   &160\p9        &1618\p101       &45     \\
G0.32-0.19     &Compact       &17 47 07.22   &-28 45 58.7   &68\p10\tablenotemark{\dag}         &$...$ & $...$    \\
G0.31-0.20     &Extended      &17 47 09.76   &-28 46 23.2   &34\p10     &64\p26  &13\\
G0.21-0.00     &Compact       &17 46 07.40   &-28 45 27.9   &182\p10        &447\p32     &18     \\
G0.17+0.15     &Extended      &17 45 26.13   &-28 42 43.9   &63\p10     &124\p27     &15   \\
G359.87+0.18   &Compact       &17 44 37.08   &-28 57 06.5   &101\p10\tablenotemark{\dag}        & $...$  &  $...$  \\
G359.87-0.09   &Compact       &17 45 38.34   &-29 05 44.1   &191\p10        &\phn446\p31     &17    \\
G359.73-0.04   &Compact       &17 45 06.27   &-29 11 19.9   &122\p10        &\phn634\p57     &31    \\
G359.72-0.04   &Compact       &17 45 05.06   &-29 11 45.7   &151\p10        &722\p53     &29     \\
G359.65-0.06   &Extended      &17 44 59.41   &-29 16 05.0   &\phn98\p9     &346\p42     &24    \\
G359.65-0.08   &Extended      &17 45 05.03   &-29 16 45.2   &143\p9        &1295\p92        &43     \\
G359.47-0.17   &Extended      &17 45 00.88   &-29 28 47.6   &72\p9     &385\p58     &31   \\
G359.36+0.00   &Compact       &17 44 06.00   &-29 28 43.5   &26\p9     &72\p35  &20    \\
G359.31-0.06   &Compact       &17 44 10.44   &-29 33 25.7   &29\p9     &121\p48     &27    \\
G359.28-0.26   &Extended      &17 44 55.66   &-29 41 10.6   &128\p9        &884\p72     &36   \\
\enddata
\tablenotetext{\dag}{Source is unresolved and we list peak intensity (mJy \beam)}
\end{deluxetable}

\begin{deluxetable}{llccccc}
\tabletypesize{\small}
\tablecaption{1.4 GHz Continuum Source Counterparts}
\tablehead{\colhead{Source} &
\colhead{Source} &
\colhead{RA} &
\colhead{DEC} &
\colhead{$S_{1.4-Lang}$} &
\colhead{$S_{1.4-FYZ}$} &
\colhead{$S_{5.0-Becker}$}\\
\colhead{Name} &
\colhead{Type} &
\colhead{(h m s)} &
\colhead{(d \am \as)} &
\colhead{(mJy)} &
\colhead{(mJy)} &
\colhead{(mJy)}}
\tablecolumns{7}
\startdata
G0.60-0.20     &Compact       &17 47 50.23   &-28 31 25.3   &31\p10\tablenotemark{\dag}     &11\p3       &$...$\\
G0.53+0.18     &Compact       &17 46 10.01   &-28 23 24.7   &504\p24                        &2364\p25    &303\p30\\
G0.48-0.10     &Compact       &17 47 08.20   &-28 34 48.3   &65\p22                         &43\p9       &$...$\\
G0.47-0.10     &Compact       &17 47 06.00   &-28 35 11.2   &898\p26                        &34\p7       &$...$\\
G0.43-0.06     &Compact       &17 46 51.39   &-28 36 09.1   &72\p21                         &55\p6       &$...$\\
G0.38+0.02     &Extended      &17 46 27.79   &-28 36 03.6   &1545\p43                       &820\p80     &1417\p425\\
G0.35-0.03     &Compact       &17 46 32.68   &-28 39 14.1   &178\p28                        &62\p28      &80\p24\\
G0.33-0.01     &Extended      &17 46 27.65   &-28 39 25.0   &1618\p101                      &910\p170    &134\p40\\
G0.32-0.19     &Compact       &17 47 07.22   &-28 45 58.7   &68\p10\tablenotemark{\dag}     &86\p5       &$...$\\
G0.31-0.20     &Extended      &17 47 09.76   &-28 46 23.2   &64\p26                         &61\p5       &$...$\\
G0.21-0.00     &Compact       &17 46 07.40   &-28 45 27.9   &447\p32                        &212\p34   &154\p46\\
G0.17+0.15     &Extended      &17 45 26.13   &-28 42 43.9   &124\p27                        &128\p2      &$...$\\
G359.87+0.18   &Compact       &17 44 37.08   &-28 57 06.5   &101\p10\tablenotemark{\dag}    &154\p2      &37\\
G359.87-0.09   &Compact       &17 45 38.34   &-29 05 44.1   &446\p31                        &130\p7      &$...$\\
G359.47-0.17   &Extended      &17 45 00.88   &-29 28 47.6   &385\p58                        &189\p15     &$...$\\
G359.36+0.00   &Compact      &17 44 06.00   &-29 28 43.5   &72\p35                         &29\p3       &4\p1\\
\enddata
\tablenotetext{\dag}{Source is unresolved and we list peak intensity (mJy \beam)}
\end{deluxetable}

\begin{deluxetable}{cccccc}
\tabletypesize{\small}
\tablecaption{Gaussian fits to $\tau_{HI}$}
\tablehead{\colhead{Source} &
\colhead{Source} &
\colhead{Velocity} &
\colhead{FWHM ($\Delta$V)} &
\colhead{Peak $\tau_{HI}$} &
\colhead{Additional Notes}\\
\colhead{Number} &
\colhead{Name}&
\colhead{(\kms)}& \colhead{(\kms)}&
\colhead{} & \colhead{}}
\tablecolumns{6}
\startdata
1 &SNR 0.9+0.1		&-40.4\p0.1	&3.9\p0.4	&4.4\p0.3	&saturated\\
$...$ &$...$		&-24.7\p0.2	&5.0\p0.4	&4.4\p0.3	&saturated\\
$...$ &$...$		&65.8\p0.4	&4.7\p1.1	&1.5\p0.3	&\\
2 &OF38	 (LR67)       	&-39.5\p0.3	&2.9\p1.1	&4.9\p1.8	&saturated\\
$...$ &$...$		&12.0\p0.5	&16.9\p1.3	&2.8\p0.2	&\\
$...$ &$...$		&53.9\p0.6	&14.0\p1.5	&1.9\p0.2	&\\
$...$ &$...$		&-14.2\p0.7\tablenotemark{\dag}	&6.1\p1.6	&1.1\p0.2	&\\
$...$ &$...$		&-24.4\p0.6\tablenotemark{\dag}	&5.7\p1.6	&1.1\p0.2	&\\
3 &SgrB2	        &61.0\p0.1	&21.5\p0.4	&4.2\p0.1	&+65 \kms~molecular cloud\\
$...$ &$...$		&-40.9\p0.1	&3.5\p0.3	&2.4\p0.2	&\\
$...$ &$...$		&-19.7\p1.7	&19.0\p5.1	&2.0\p0.4	&\\
$...$ &$...$		&17.8\p0.2	&5.8\p0.4	&1.9\p0.1	&\\
$...$ &$...$		&-49.6\p3.2	&22.0\p6.6	&0.4\p0.1	&3-kpc arm\\
4 &G0.6-0.0 (LR60)		&51.1\p0.2	&16.4\p0.4	&4.3\p0.1	&saturated\\
$...$ &$...$	   	&-41.6\p0.1	&5.4\p0.3	&3.8\p0.2	&saturated\\
$...$ &$...$		&18.6\p0.4	&5.3\p0.9	&1.4\p0.2	&\\
$...$ &$...$	    	&-58.9\p0.3	&4.5\p0.7	&1.3\p0.2	&3-kpc arm\\
$...$ &$...$		&-24.0\p1.4	&15.4\p3.5	&0.8\p0.2	&\\
5 &SgrB1 New		&-42.2\p0.3	&5.1\p0.8	&2.0\p0.3	&\\
$...$ &$...$		&-58.9\p0.4 	&3.8\p0.8	&1.5\p0.3	&3-kpc arm\\
$...$ &$...$		&19.3\p0.7	&9.5\p1.6	&1.4\p0.2	&\\
$...$ &$...$		&43.9\p1.3	&10.7\p3.0	&0.7\p0.2	&\\
$...$ &$...$		&-22.7\p1.0\tablenotemark{\dag}	&9.7\p2.5	&0.7\p0.2	&\\
6 &G0.53+0.18		&-28.2\p0.4	&4.0\p0.7	&1.2\p0.2	&\\
$...$ &$...$		&-56.6\p0.7	&4.0\p1.2	&0.7\p0.2	&3-kpc arm\\
7 &G0.38+0.02 (LR54) 	&-54.7\p0.1	&4.7\p0.2	&3.9\p0.1	&3-kpc arm,saturated\\
$...$ &$...$		&19.3\p0.1	&4.6\p0.3	&3.5\p0.2	&saturated\\
$...$ &$...$		&29.6	&14.3	&1.2	&hand fit\\
$...$ &$...$		&-25.8	&10.0	&0.6	&hand fit\\
$...$ &$...$		&46.1\p1.1	&6.7\p2.6	&0.4\p0.1	&\\
8 &G0.34-0.01		&19.0\p0.1	&4.7\p0.4	&4.3\p0.2	&\\
$...$ &$...$		&-54.7\p0.1	&4.2\p0.3	&4.0\p0.2	&3-kpc arm\\
$...$ &$...$		&34.9\p1.8	&18.4\p4.9	&0.6\p0.1	&\\
9 &SNR 0.3+0.0		&24.3\p0.3	&4.6\p0.8	&0.7\p0.1	&\\
$...$ &$...$		&-55.4\p0.3	&3.1\p6.0	&0.6\p0.9	&3-kpc arm\\
$...$ &$...$		&-26.6\p0.8	&15.4\p2.3	&0.6\p0.1	&\\
$...$ &$...$		&-49.3\p2.1	&15.8\p3.4	&0.4\p0.1	&\\
10  &G0.32-0.19		&15.4\p0.1	&6.5\p0.3	&4.8\p0.2	&\\
$...$ &$...$		&50.5\p0.7	&7.3\p1.7	&1.0\p0.2	&\\
$...$ &$...$		&66.8\p2.0	&10.6\p4.9	&0.4\p0.2	&\\
11 &G0.31-0.20		&14.2\p0.2	&3.7\p0.8	&5.7\p1.0	&\\
$...$ &$...$		&43.3\p0.3	&6.9\p0.8	&4.3\p0.4	&\\
$...$ &$...$		&72.5\p0.4	&4.7\p1.0	&2.7\p0.5	&\\
$...$ &$...$		&120.7\p1.3	&3.5\p2.2	&1.0\p0.6	&\\
12 &Radio Arc		&-53.7\p0.3	&7.4\p0.6	&1.3\p0.1	&3-kpc arm\\
$...$ &$...$		&-31.4\p0.7	&10.1\p1.7	&0.6\p0.1	&\\
13 &Pistol		&11.3\p0.2	&6.0\p0.5	&4.0\p0.3	&saturated\\
$...$ &$...$		&-53.1\p0.7	&6.9\p1.6	&1.2\p0.2	&3-kpc arm\\
$...$ &$...$		&-32.0\p1.0	&5.9\p2.5	&0.7\p0.3	&\\
14 &Sickle		&-31.7\p0.3	&6.1\p0.8	&0.8\p0.1	&\\
$...$ &$...$		&-51.7\p0.5	&13.8\p1.2	&0.8\p0.1	&3-kpc arm\\
15 &Arches-E1		&-54.8\p0.4	&8.5\p0.9	&0.8\p0.1	&3-kpc arm\\
$...$ &$...$		&-29.7\p0.5	&9.5\p1.2	&0.7\p0.1	&$-$30 \kms~molecular cloud\\
$..$ &$...$		&-41.8\p0.6	&5.6\p1.6	&0.4\p0.1	&\\
16 &Arches-E2		&-55.8\p0.3	&6.2\p0.7	&0.7\p0.1	&3-kpc arm\\
$...$ &$...$		&-27.6\p0.6	&18.3\p1.6	&0.6\p0.1	&$-$30 \kms~molecular cloud\\
17 &G0.10+0.02		&-54.3\p0.5	&7.0\p1.2	&1.5\p0.2	&3-kpc arm\\	
$...$ &$...$		&-28.7\p1.1	&8.8\p2.6	&0.8\p0.2	&\\
18 &Arches-West		&-25.9\p0.3	&3.7\p0.6	&1.2\p0.2	&$-$30 \kms~molecular cloud\\
$...$ &$...$		&-35.2\p1.7	&24.4\p4.2	&0.7\p0.1	&$-$30 \kms~molecular cloud\\
$...$ &$...$		&-58.7\p0.8	&8.1\p1.9	&0.6\p0.1	&3-kpc arm\\
$...$ &$...$		&22.1\p0.9	&8.2\p2.2	&0.5\p0.1	&\\
$...$ &$...$		&65.5\p3.8	&17.9\p8.8	&0.3\p0.1	&\\
$...$ &$...$		&43.9\p5.1	&16.0\p12.2	&0.2\p0.1	&\\
19 &G0.17+0.15		&-22.8\p0.6	&3.7\p1.1	&1.4\p0.4	&\\
$...$ &$...$		&65.8\p2.4	&16.3\p6.1	&0.7\p0.2	&\\
$...$ &$...$		&-77.9\p2.1	&3.6\p4.5	&0.4\p0.4	&\\		
$...$ &$...$		&-143.1\p2.5	&6.0\p6.1	&0.4\p0.3	&\\
$...$ &$...$		&20.2\p5.6	&19.1\p14.6	&0.4\p0.2	&\\
20 &H5		&23.9\p0.1	&4.8\p0.3	&1.9\p0.1	&\\
$...$ &$...$		&-55.8\p0.6	&07\p1.4	&0.6\p0.1	&\\
$...$ &$...$		&-28.0\p0.5	&9.0\p1.4	&0.6\p0.1	&$-$30 \kms~molecular cloud\\
$...$ &$...$		&-41.1\p1.2	&5.8\p2.8	&0.2\p0.1	&\\
		&               &               &               &\\
21 &H3		&-58.9\p0.3	&4.4\p0.6	&1.1\p0.1	&\\
$...$ &$...$		&20.9\p0.5	&5.1\p1.2	&0.7\p0.1	&\\
$...$ &$...$		&-21.5\p0.8	&8.2\p2.1	&0.6\p0.1	&\\
$...$ &$...$		&-35.9\p1.6	&11.1\p4.3	&0.4\p0.1	&$-$30 \kms~molecular cloud\\
$...$ &$...$		&-75.0\p1.2	&5.1\p2.8	&0.3\p0.1	&\\
22 &H2		&20.7\p0.3	&3.4\p0.4	&1.4\p0.1	&\\
$...$ &$...$		&-56.1\p0.3	&7.1\p0.8	&1.2\p0.1	&3-kpc arm\\
$...$ &$...$		&-17.9	&5.6	&0.4	&hand fit\\
$...$ &$...$		&-37.0	&8.4	&0.4	&hand fit, $-$30 \kms~molecular cloud\\
23 &H1		&-55.5\p0.6	&9.9\p1.4	&1.0\p0.1	&\\
$...$ &$...$		&22.0\p0.6	&5.5\p1.4	&0.6\p0.1	&\\
$...$ &$...$		&-30.6	&17.1	&0.4	&hand fit,$-$30 \kms~molecular cloud\\
$...$ &$...$		&-20.4\p1.4	&4.4\p2.9	&0.3\p0.2	&\\
24 &SgrA East		&-53.3\p0.4	&7.2\p0.9	&1.2\p0.1	&3-kpc arm\\
$...$ &$...$		&47.1\p1.8	&26.9\p4.7	&0.6\p0.1	&+50 \kms~molecular cloud\\
$...$ &$...$		&18.8\p2.0	&14.0\p5.6	&0.5\p0.1	&\\
$...$ &$...$		&-29.0\p2.2	&16.3\p5.4	&0.3\p0.1	&\\
25 &SgrA West		&-53.4\p0.5	&8.9\p1.1	&1.0\p0.1	&3-kpc arm\\
$...$ &$...$		&-29.6\p1.6	&16.8\p4.0	&0.4\p0.1	&\\
$...$ &$...$		&20.5\p1.2	&9.8\p4.1	&0.4\p0.2	&\\
$...$ &$...$		&40.0\p10.9	&42.3\p22.3	&0.3\p0.1	&\\
26 &Southern Thread	&-58.9	&4.8	&1.6	&hand fit,3-kpc arm\\
$...$ &$...$		&-24.9	&11.4	&1.0	&hand fit\\
$...$ &$...$		&-42.0	&13.2	&0.6	&hand fit\\
27 &G359.87+0.18 (LR44) 	&11.7\p0.2	&2.9\p2.2	&5.5\p4.4	&\\
$...$ &$...$		&-11.7\p0.2	&5.1\p0.5	&4.5\p0.3	&\\
$...$ &$...$		&-25.0\p0.5	&3.7\p1.5	&1.4\p0.4	&\\
$...$ &$...$		&26.8\p2.4	&5.3\p6.0	&0.3\p0.3	&\\
28 &G359.79+0.17	&-26.9\p0.1	&5.0\p0.3	&4.7\p0.3	&saturated\\
$...$ &$...$		&25.4\p2.6	&2.1\p2.9	&3.5\p0.2	&\\
$...$ &$...$		&64.3\p1.7	&18.1\p4.2	&0.7\p0.1	&\\
$...$ &$...$		&-60.6	&6.8	&0.3	&hand fit,3-kpc arm\\
29 &G359.73-0.03 (LR40)		&-26	&4.1	&2.3	&hand fit\\
$...$ &$...$		&15.7	&2.9	&1.6	&hand fit\\
$...$ &$...$		&-54.1	&5.3	&1.1	&hand fit,3-kpc arm\\
30 &G359.72-0.04		&-25.8	&7.4	&1.2	&hand fit\\
$...$ &$...$		&-54.1	&7.1	&1.0	&hand fit\\
31 &G359.54+0.18		&-26.9\p0.1	&3.1\p0.5	&5.5\p0.8	&saturated\\
$...$ &$...$		&12.7\p0.3	&6.3\p0.8	&1.3\p0.1	&\\
$...$ &$...$		&-40.3\p1.9	&23.3\p4.4	&0.5\p0.1	&\\
$...$ &$...$		&56.5	&4.2\p1.8	&0.2\p0.1	&\\
32 &G359.65-0.07		&-20.3\p0.2	&3.2\p0.3	&2.9\p0.2	&\\
$...$ &$...$		&-52.8\p0.2	&5.0\p0.4	&2.2\p0.3	&3-kpc arm\\
$...$ &$...$		&14.9\p0.3	&3.0\p0.6	&1.8\p0.3	&\\
$...$ &$...$		&-13.4\p0.2	&5.2\p0.6	&1.7\p0.1	&\\
$...$ &$...$		&-30.1\p0.3	&4.6\p0.8	&1.1\p0.2	&\\
$...$ &$...$		&-142.5\p2.1	&14.3\p5.5	&0.3\p0.1	&\\
33 &G359.65-0.06 (LR37)	&-53.5\p0.1	&3.6\p0.2	&3.8\p0.2 &3-kpc arm\\
$...$ &$...$		&-18.1\p0.3	&3.1\p0.3	&1.4\p0.1	&\\
$...$ &$...$		&13.4\p0.2	&4.5\p0.5	&1.4\p0.2	&\\
$...$ &$...$		&-42.8\p0.6	&14.1\p2.1	&1.1\p0.1	&\\
$...$ &$...$		&-29.6\p0.5	&5.2\p1.3	&0.7\p0.1	&\\
34 &G359.65-0.08 (LR36)	&-16.6\p0.1	&4.8\p0.2	&4.7\p0.1\\
$...$ &$...$		&-53.7\p0.2	&6.4\p0.4	&2.0\p0.1	&3-kpc arm\\
$...$ &$...$		&15.9\p0.3	&3.6\p0.5	&1.1\p0.1	&\\
$...$ &$...$		&-28.3\p0.6	&7.3\p1.4	&0.6\p0.1	&\\
35 &FIR-4		&-28.8\p1.5	&2.3\p2.1	&4.0\p0.2	&saturated\\
$...$ &$...$		&-133.7\p1.4	&15.0\p3.3	&1.4\p0.3	&\\
$...$ &$...$		&24.4\p1.0	&4.6\p3.1	&1.0\p0.5	&\\
$...$ &$...$		&-54.3\p5.1	&12.0\p12.2	&0.4\p0.3	&3-kpc arm\\
36 &SgrC Fil		&12.3\p0.2	&7.1\p0.5	&3.4\p0.2	&\\
$...$ &$...$		&-27.9\p1.1	&10.5\p2.8	&0.6\p0.1	&\\
$...$ &$...$		&25.0\p1.2	&10.4\p3.1	&0.7\p0.1	&\\
$...$ &$...$		&-54.0\p2.3	&4.8\p3.8	&0.7\p0.2	&3-kpc arm\\
$...$ &$...$		&-59.9\p2.9	&5.2\p6.2	&0.5\p0.2	&$-$65 \kms~molecular cloud\\
$...$ &$...$		&-104.8\p5.5	&14.1\p15.7	&0.2\p0.1	&\\
37 &SgrC HII		&14.3\p0.1	&4.6\p0.4	&3.1\p0.2	&\\
$...$ &$...$		&-53.7\p0.7	&4.0\p1.2	&1.3\p0.2	&3-kpc arm\\
$...$ &$...$		&-27.2\p0.6	&6.1\p1.4	&0.8\p0.2	&\\
$...$ &$...$		&-114.0\p4.8	&13.4\p11.9	&0.2\p0.2	&\\
$...$ &$...$		&-133.4\p9.1	&27.1\p19.1	&0.3\p0.1	&\\
$...$ &$...$		&-60.1\p0.7	&4.8\p2.3	&1.0\p0.2	&$-$65 \kms~molecular cloud\\
38 &S-SgrC (LR21)	&14.4\p0.2      &6.6\p0.4       &4.2\p0.2       &saturated\\
$...$ &$...$		&-53.4\p0.4     &3.9\p0.9    	&1.7\p0.3       &3-kpc arm\\
$...$ &$...$		&-28.4\p0.6     &3.3\p0.8       &1.3\p0.3       &\\
$...$ &$...$		&30.2\p2.1      &4.9\p5.1       &0.3\p0.3       &\\
$...$ &$...$		&-17.5\p0.4\tablenotemark{\dag}	&3.5\p0.7	&1.2\p0.2	&\\
39 &G359.47-0.17 (LR26)	&-14.0\p0.2	&6.0\p0.6	&4.0\p0.3	&saturated\\
$...$ &$...$		&-33.3	&28.5	&0.9	&hand fit\\
$...$ &$...$		&-60.6\p2.4	&14.8\p5.3	&0.7\p0.2	&3-kpc arm\\
$...$ &$...$		&-117.0\p5.0	&21.4\p13.1	&0.3\p0.1	&\\
$...$ &$...$		&-140.9\p3.6	&8.3\p9.1	&0.3\p0.2	&\\
40 &G359.28-0.26 (LR14)	&13.8\p0.1	&4.9\p0.4	&2.8\p0.2	&\\
$...$ &$...$		&-38.7	&7.3	&0.4	&hand fit\\
\enddata
\end{deluxetable}

\begin{deluxetable}{cccccccc}
\tablecaption{Properties of Negative Velocity HI Clouds}
\tablewidth{0pt}
\tablehead{
\colhead{Cloud} &
\colhead{Source} &
\colhead{$\upsilon_{HI}$} &
\colhead{$\upsilon_{rms}$} &
\colhead{$\Delta\upsilon_{HI}$} &
\colhead{$\Delta\upsilon_{rms}$} &
\colhead{$\tau_{HI}$} &
\colhead{$\tau_{rms}$}\\
\colhead{} &
\colhead{Numbers} &
\colhead{(km s$^{-1}$)} &
\colhead{(km s$^{-1}$)} &
\colhead{(km s$^{-1}$)} &
\colhead{(km s$^{-1}$)} &
\colhead{} &
\colhead{}}
\tablecolumns{8}
\startdata
A & 3-9,12-18,21-26,28-30,32-9 &-56.9 &0.9 &6.32 &2.2 &1.22 &0.2\\
B & 1-5,15,20,26,31,33,40      &-41.0 &0.6 &8.19 &1.5 &1.88 &0.4\\ 
C & 1,2,4,6,7,9,12-8,20,23-38  &-27.7 &0.8 &7.80 &1.9 &1.32 &0.2\\ 
D & 3,5,19,21-23,32,33,38      &-20.1 &1.0 &7.25 &2.5 &1.15 &0.2\\ 
\enddata
\end{deluxetable}

\begin{deluxetable}{lccccccc}
\tablecaption{The ISM in the SgrB 1,2 and G0.06-0.0 regions}
\tabletypesize{\small}
\tablewidth{0pt}
\tablehead{
\colhead{Region} &
\colhead{$\upsilon_{HI}$} &
\colhead{$\tau_{HI}$} &
\colhead{$\Delta\upsilon_{HI}$} &
\colhead{$\upsilon_{H_{2}CO}$} &
\colhead{N$_{HI}$/N$_{H_2}$} &
\colhead{$\upsilon_{HII}$} &
\colhead{RRL} \\
\colhead{Name}&
\colhead{(km s$^{-1}$)} &
\colhead{opacity} &
\colhead{(km s$^{-1}$)} &
\colhead{(km s$^{-1}$)} &
\colhead{Ratio} &
\colhead{(km s$^{-1}$)} &
\colhead{Region}}
\tablecolumns{8}
\startdata
SgrB2(N)$-$18 &67.4\p0.3 & 1.8 & 22.4 & 63.4\p0.1 & 0.02 & 70.5\p1.0 & 3\\
SgrB2(R)$-$20 &53.7\p0.6 & 1.2 & 20.6 & 51.3\p0.2 & 0.03 & 51.0\p1.0 & 1\\
SgrB2(M)$-$15 &63.5\p0.3 & 2.0 & 22.6 & 65.3\p0.1 & 0.03 & 61.0\p1.0 & 4\\
SgrB2(U)$-$17 &62.4\p0.3 & 2.0 & 17.8 & 61.2\p0.1 & 0.03 & 64.0\p1.0 & 10\\
SgrB2(V)$-$13 &64.1\p0.6 & 1.1 & 14.1 & 65.1\p0.1 & 0.03 & 99.0\p1.0 & 9 \\
SgrB2(S)$-$12 &51.0\p0.2 & 3.5 & 14.6 & 53.9\p0.5 & 0.05 & 59.0\p1.0 & 7 \\
G0.6-0.0(9)  &51.7\p0.4 & 1.4 & 19.2 & 51.6\p0.2& 0.18 & 55.0\p1.0 & 1\tablenotemark{*}\\
SgrB1$-$W(3) & 42.9\p0.8 & 0.4 & 8.6 & $\sim$40 & $-$ & 41-47 & 11,12\tablenotemark{*}\\
SgrB1$-$E(4) & 47.2\p0.5 & 1.5 & 16.2 & 42-54 & 0.10 & 53-63 & 3-6\tablenotemark{*}\\
\enddata
\tablenotetext{*}{Names from H110$\alpha$ study by Mehringer et al. (1992).}
\end{deluxetable}

\begin{deluxetable}{lccccc}
\tablecaption{The ISM in the Radio Arc region}
\tabletypesize{\small}
\tablewidth{0pt}
\tablehead{

\colhead{Region} &
\colhead{$\upsilon_{HI}$} &
\colhead{$\tau_{HI}$} &
\colhead{$\Delta\upsilon_{HI}$} &
\colhead{$\upsilon_{CS}$}&
\colhead{$\upsilon_{HII}$} \\

\colhead{Name}&
\colhead{(km s$^{-1}$)} &
\colhead{opacity} &
\colhead{(km s$^{-1}$)} &
\colhead{(km s$^{-1}$)} &
\colhead{(km s$^{-1}$)}}

\tablecolumns{6}
\startdata
Arches-E1&+13.5\p0.2 & 1.2 & 4.1\p0.4 & $-$25 & $-$18.0\p0.4\\
         &$-$28.4\p0.7 & 0.4 & 9.5\p1.7 & $-$ & $-$\\
         &$-$53.9\p0.5 & 0.4 & 4.1\p0.4 & $-$ & $-$\\

Arches-E2&$-$29.2\p0.9 & 0.4 & 20.0\p2.2& $-$20 & $-$33.0\p0.7\\
         &$-$55.9\p0.4 & 0.5 & 6.5\p0.9& $-$ & $-$\\
        
Arches-G0.12+0.02&$-$28.7\p1.1 & 0.8 & 8.8\p2.6 & $-$25 & $-$28.1\p0.6 \\
                 &$-$54.4\p0.6 & 0.9 & 7.4\p0.9& $-$ & $-$\\

Arches-West&+20.1\p0.6 & 0.4 & 11.6\p1.7&$-$20 & $-$26.4\p0.4\\
           &$-$27.5\p0.3 & 0.9 & 10.3\p0.8 & $-$ & $-$\\
           &$-$40.7\p0.4 & 0.6 & 7.8\p1.3& $-$ & $-$\\
           &$-$55.8\p1.7 & 0.3 & 21.4\p4.0& $-$ & $-$\\

Arches-W1 &+23.7\p0.5 & 0.3 & 3.8\p1.0 & $-$40 & $-$43.6\p0.4\\
           &$-$31.9\p1.2 & 0.3 & 24.1\p3.5& $-$ & $-$\\
           &$-$57.6\p0.3 & 0.6 & 6.8\p0.8& $-$ & $-$\\

Arches-W2&+23.5\p0.6 & 0.4 & 4.5\p1.4 & $-$35 & $-$27.0\p0.4\\
           &$-$32.7\p0.7 & 0.9 & 19.2\p1.7& $-$ & $-$ \\
           &$-$59.1\p0.7 & 0.4 & 7.9\p1.7& $-$ & $-$ \\

Sickle-1 &$-$31.3\p0.4 & 0.5 & 5.5\p1.0 & $-$ & $-$17.3\p2.1\\
         & $-$51.1\p0.6 & 0.5 & 16.2\p1.5 & $-$ & 35.7\p1.8\\

Sickle-2&$-$31.7\p0.2 & 0.7 & 4.1\p0.5 & +21.4\p0.2 & 31.2\p0.7\\
        &$-$52.2\p0.4 & 0.5 & 10.9\p1.1 & $-$ & $-$\\

Sickle-3&$-$32.8\p0.2 & 0.7 & 6.1\p0.5 & +25.4\p0.2 & 53.0\p0.7\\
        &$-$53.0\p0.6 & 0.4 & 15.4\p1.6 & $-$ & 5.9\p0.1\\

Pistol Nebula & $-$32.3\p0.6& 0.4 & 7.0\p1.5 & $-$ & 111.0\p1.1\\
              & $-$53.2\p0.4& 0.7 & 7.4\p0.8 & $-$ & 111.0\p1.1\\

H1  & +22.2\p0.3& 0.4 & 5.0\p0.8 & $-$ & $-$39\p6\\
    & $-$29.6\p0.9& 0.3 & 8.0\p1.0 & $-$ & $-$\\
    & $-$55.7\p0.3& 0.6 & 10.7\p0.7 & $-$ & $-$\\
    
H2  & +20.8\p0.3& 0.7 & 3.6\p0.5 & $-$ & $-$58\p6\\
    & $-$55.9\p0.4 & 0.8 & 8.1\p0.8 & $-$ & $-$\\

H3  & +20.1\p0.6& 0.3 & 6.4\p1.4 & $-$ & $-$42\p7\\
    & $-$21.1\p0.7& 0.3 & 7.9\p2.3 & $-$ & $-$ \\
    & $-$58.9\p0.3& 0.6 & 4.2\p0.6 & $-$ & $-$ \\

H5  & +23.9\p0.1& 0.9 & 5.2\p0.3 & $-$45& $-$36\p7\\
    & $-$28.3\p0.4& 0.4 & 9.1\p1.0 & $-$25 & $-$ \\
    & $-$55.1\p0.4& 0.4 & 11.8\p1.2 & $-$ & $-$ \\
\enddata

\tablenotetext{\dag}{CS velocities for the Arches and H5 are from 
Serabyn \& Guesten (1987) and trace the CS (2-1) transition, and for the
 Sickle are from Serabyn \& Guesten (1991) and trace the CS (3-2) transition.}
\tablenotetext{\ddag}{Radio recombination line velocities for the Arches from
the H92$\alpha$ measurements of Lang et al. (2001), for the Sickle from the
H92$\alpha$ study of Lang et al. (1997), and for the 'H'-regions, from
the H110$\alpha$ study by Zhao et al. (1993).}
\end{deluxetable}

\begin{figure}
\includegraphics[width=16cm]{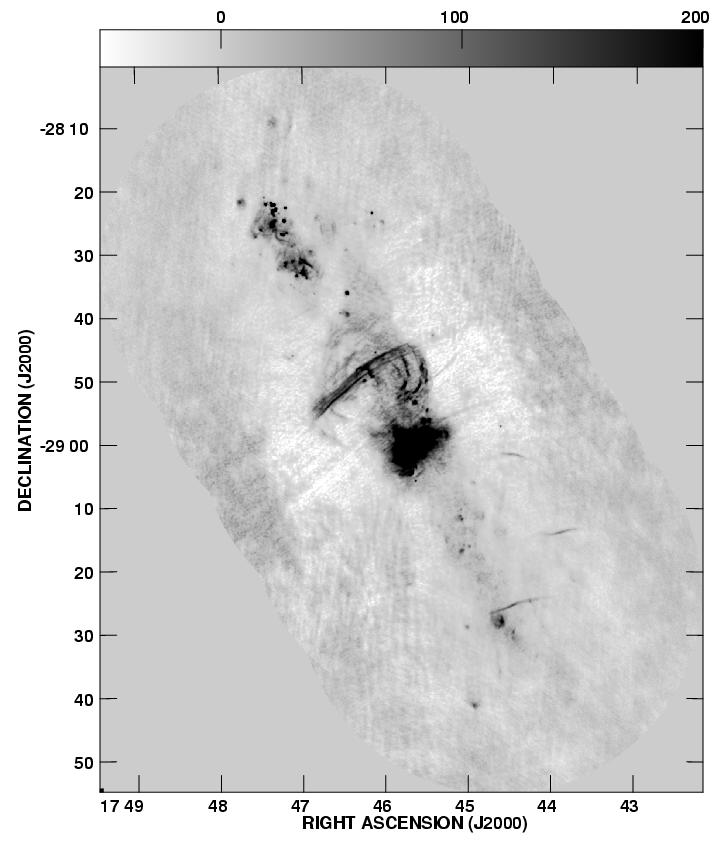}
\caption{Greyscale image representing the 1.4 GHz VLA continuum data associated with our
HI observation. The resolution in this image is 15\arcsec, and the rms level is between 3-4 mJy \beam. 
The image was assembled from approximately 16 line-free channels. 
\label{fig12}}
\end{figure}

\begin{figure}
\includegraphics[width=16cm]{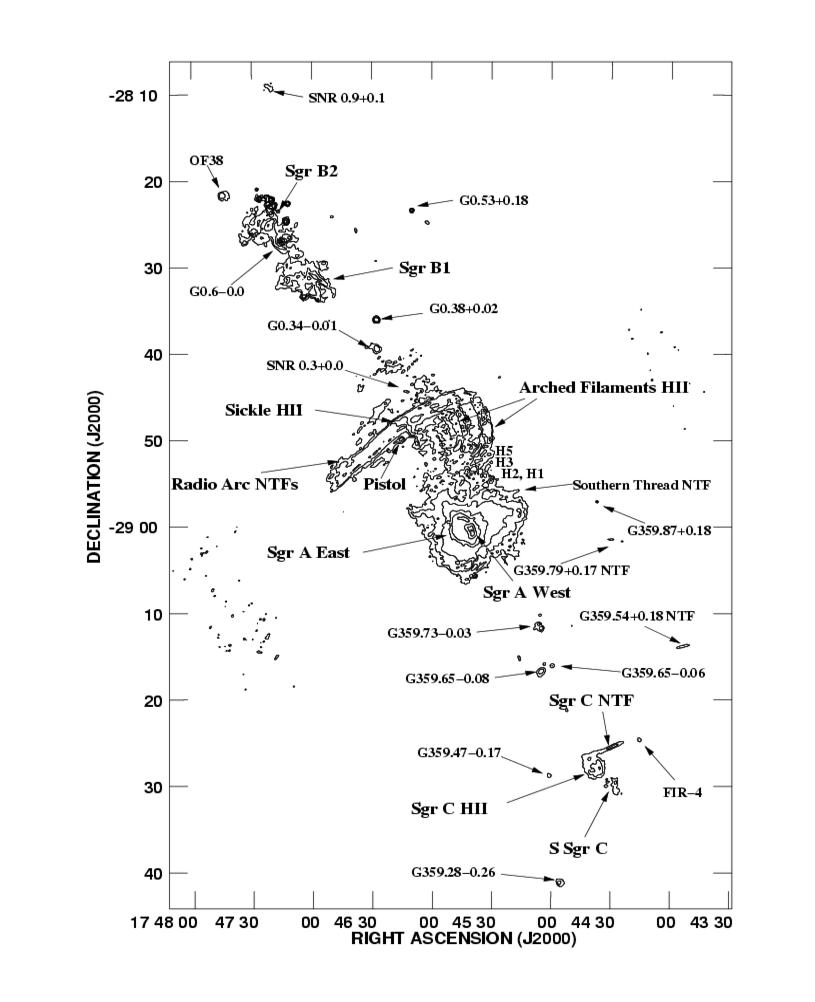}
\caption{Contours representing the 1.4 GHz VLA continuum image shown in Figure 1. Major sources
and some compact sources are labeled. The resolution is 15\arcsec. The contour levels represent
5, 10, 15, 20, 25, 30, 35, 50, 100, 300, 400, 500 times the level of 8.5 mJy beam$^{-1}$. 
\label{fig12}}
\end{figure}

\begin{figure}
\includegraphics[width=16cm]{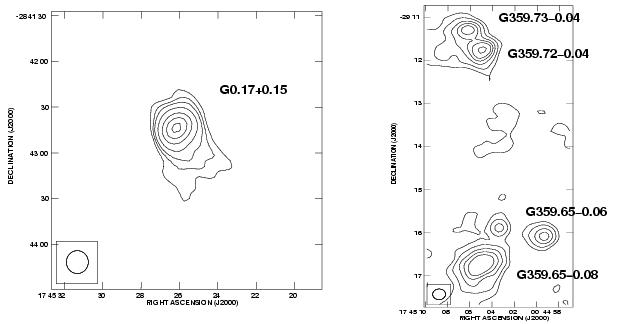}
\caption{Insets of several point sources from the 1.4 GHz continuum image (Figures 1 and 2): 
(left) G0.17+0.15, a source near the Arched Filaments and Radio Arc complex. The contour
levels represent 3, 6, 9, 15, 25, 35, 45, and 63 times 1.32 mJy beam$^{-1}$. (Right) four 
compact sources near Sgr C at negative Galactic longitudes. Contour levels represent 
4, 6, 9, 12, 15, 20, 25, 30, 35, 40, 45, 50, 55,and 63 times 7.3 mJy beam$^{-1}$.  
\label{fig12}}
\end{figure}

\begin{figure}
\begin{center}
\includegraphics[width=10cm]{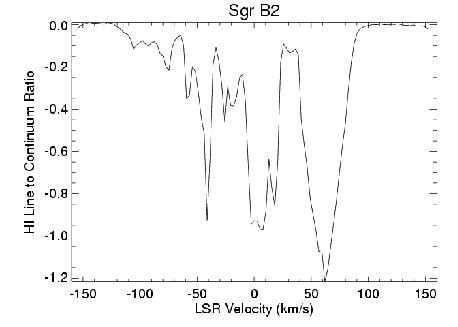} 

\includegraphics[width=10cm]{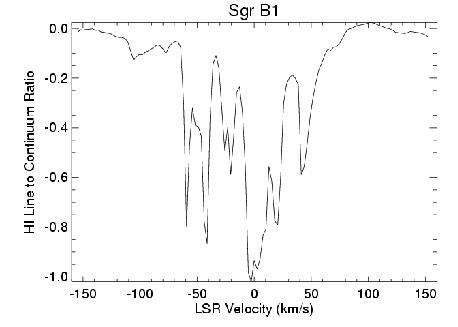} 
\caption{HI absorption spectra (in line-to-continuum ratio units) towards Sgr B2 (above) and Sgr B1 (below). 
The velocity resolution is 2.5~\kms~and the bandwidth is 1.5 MHz.} 
\label{fig19}
\end{center}
\end{figure}

\clearpage

\begin{figure}
\begin{center}
\includegraphics[width=10cm]{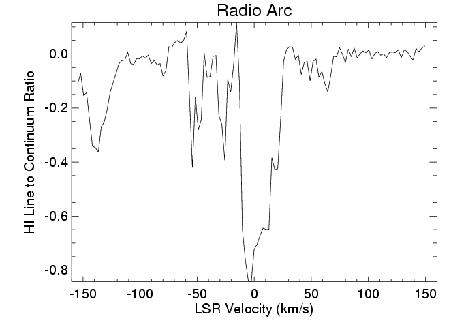} 

\includegraphics[width=10cm]{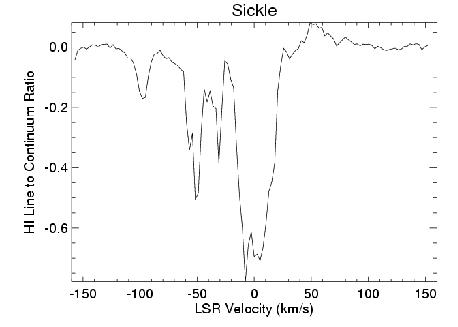} 
\caption{HI absorption spectra (in line-to-continuum ratio units) towards the Radio Arc (top) and the Sickle HII Region (bottom). 
The velocity resolution is 2.5~\kms~and the bandwidth is 1.5 MHz.} 
\label{fig19}
\end{center}
\end{figure}

\begin{figure}
\begin{center}
\includegraphics[width=10cm]{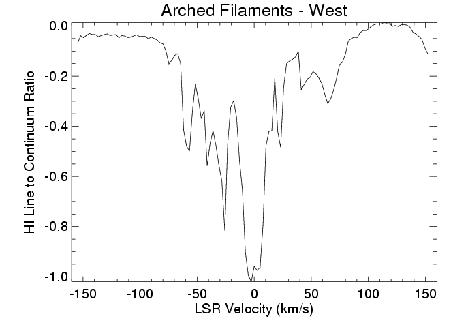} 
\caption{HI absorption spectra (in line-to-continuum ratio units) towards the Western Arched Filaments. The velocity resolution is 
2.5~\kms~and the bandwidth is 1.5 MHz.
\label{fig14}}
\end{center}
\end{figure}

\begin{figure}
\begin{center}
\includegraphics[width=10cm]{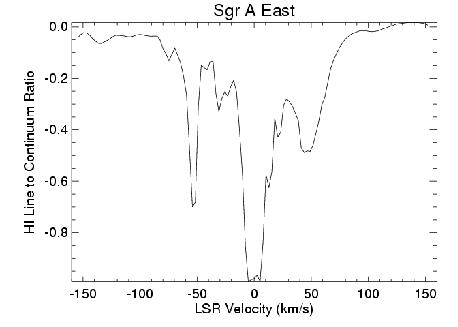} 

\includegraphics[width=10cm]{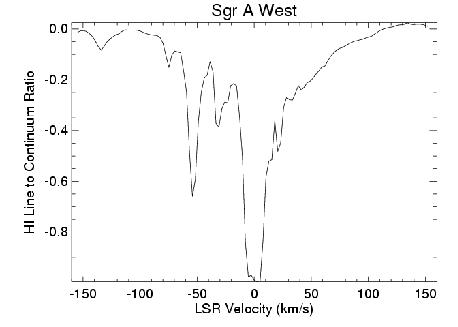} 
\caption{HI absorption spectra (in line-to-continuum ratio units) towards Sgr A East (top) and Sgr A West (bottom). 
The velocity resolution is 2.5~\kms~and the bandwidth is 1.5 MHz.} 
\label{fig19}
\end{center}
\end{figure}

\begin{figure}
\begin{center}
\includegraphics[width=10cm]{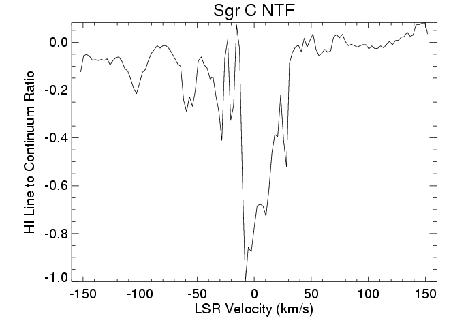} 

\includegraphics[width=10cm]{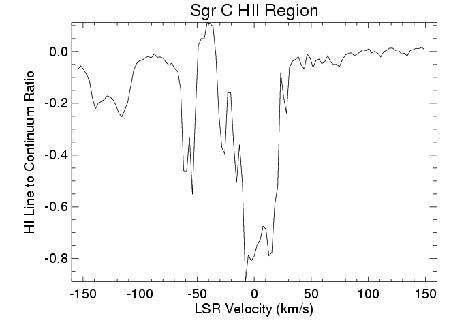} 
\caption{HI absorption spectra (in line-to-continuum ratio units) 
towards the Sgr C nonthermal filament (NTF) (top) and the Sgr C HII Region (bottom). 
The velocity resolution is 2.5~\kms and the total bandwidth is 1.5 MHz.} 
\label{fig19}
\end{center}
\end{figure}

\begin{figure}
\begin{center}
\includegraphics[width=10cm]{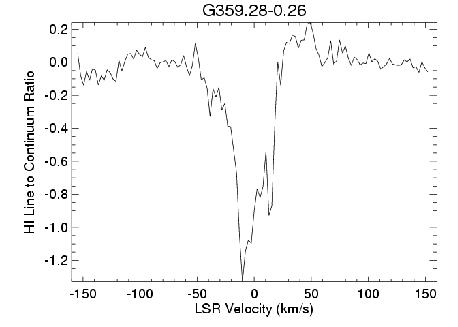} 

\includegraphics[width=10cm]{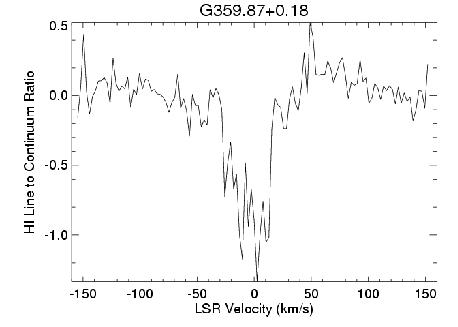} 
\caption{HI absorption spectra (in line-to-continuum ratio units) towards G359.28-0.26 (top) and G359.87+18 (bottom). 
The velocity resolution is 2.5~\kms~and the total bandwidth is 1.5 MHz.} 
\label{fig19}
\end{center}
\end{figure}

\begin{figure}
\begin{center}
\includegraphics[width=12cm]{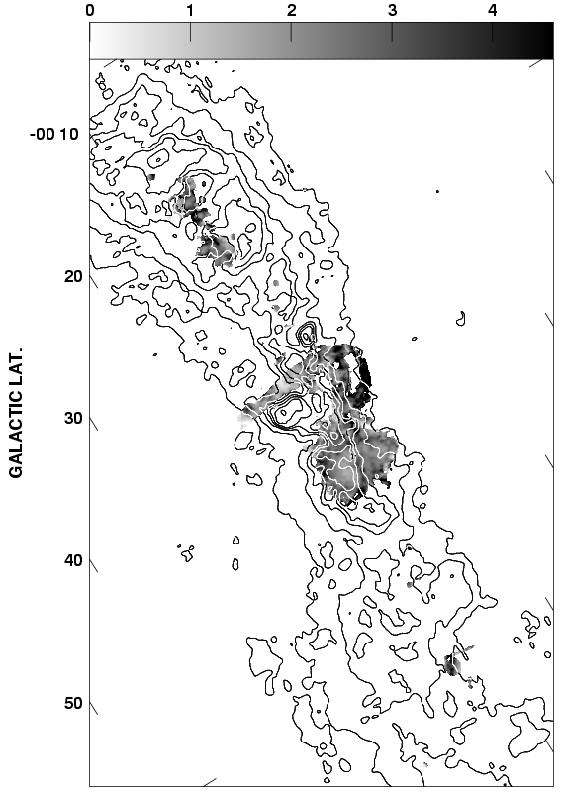} 
\caption{HI opacity at velocities of $-$20 to $-$30 \kms~toward the entire GC region shown in greyscale with values
for $\tau$=0-4. Contours represent the distribution of CO (J=1-0) emission in the range of 20-40 \kms~
from the survey of Oka et al. (1998) which has a spatial resolution of $\sim$30\arcsec. The contour levels represent 
5, 15, 25, 40, 50, 62.5 and 75 K \kms. HI absorption at these 
velocities ($-$20 to $-$30 \kms) is present toward the majority of major features in the GC (i.e., Sgr B, Radio Arc, 
Sgr A and Sgr C).
\label{fig17}}
\end{center}
\end{figure}

\begin{figure}
\begin{center}
\includegraphics[width=12cm]{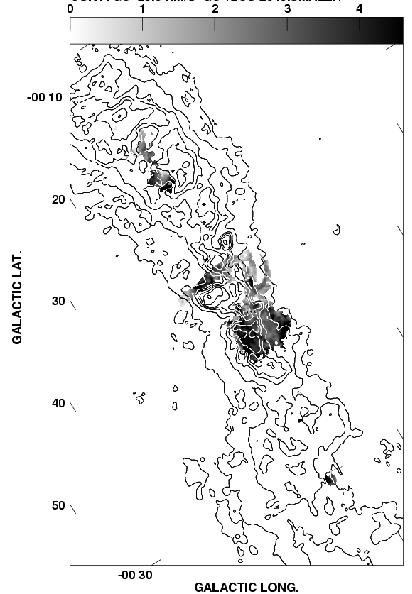} 
\caption{HI opacity at velocities of $-$50 to $-$60 \kms~toward the entire GC region shown in greyscale with a range 
for $\tau$=0-4. HI absorption in this velocity range is likely to be associated with the ``3-kpc arm'' foreground spiral
arm (v=$-$54.3 \kms). Contours represent the distribution of CO (J=1-0) emission in the range of 20-40 \kms~
from the survey of Oka et al. (1998) which has a spatial resolution of $\sim$30\arcsec. The contour levels represent 
5, 15, 25, 40, 50, 62.5 and 75 K \kms. HI absorption at these 
velocities ($-$50 to $-$60 \kms) is present toward the majority of major features in the GC (i.e., Sgr B, Radio Arc, 
Sgr A and Sgr C).
\label{fig17}}
\end{center}
\end{figure}

\begin{figure}
\includegraphics[width=16cm]{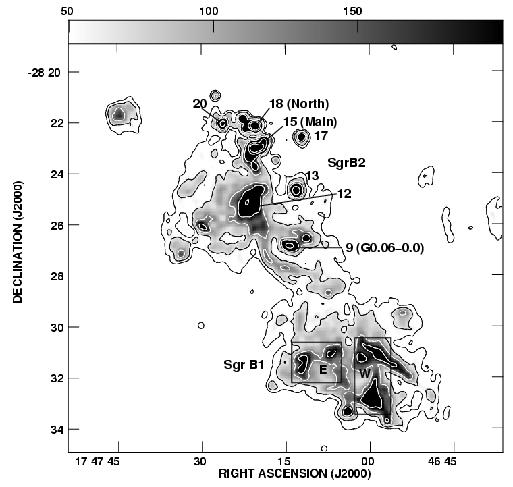}
\caption{VLA 1.4 GHz continuum image of the Sgr B complex shown in both greyscale 
and contours. The spatial resolution of this image is 15''. Small
regions for which HI opacity profiles were produced (shown in Figures
10-15) are labeled according to Mehringer et al. (1993) for Sgr B2 and
Mehringer et al. (1992) for Sgr B1. 
\label{fig17}}
\end{figure}

\begin{figure}
\includegraphics[width=16cm]{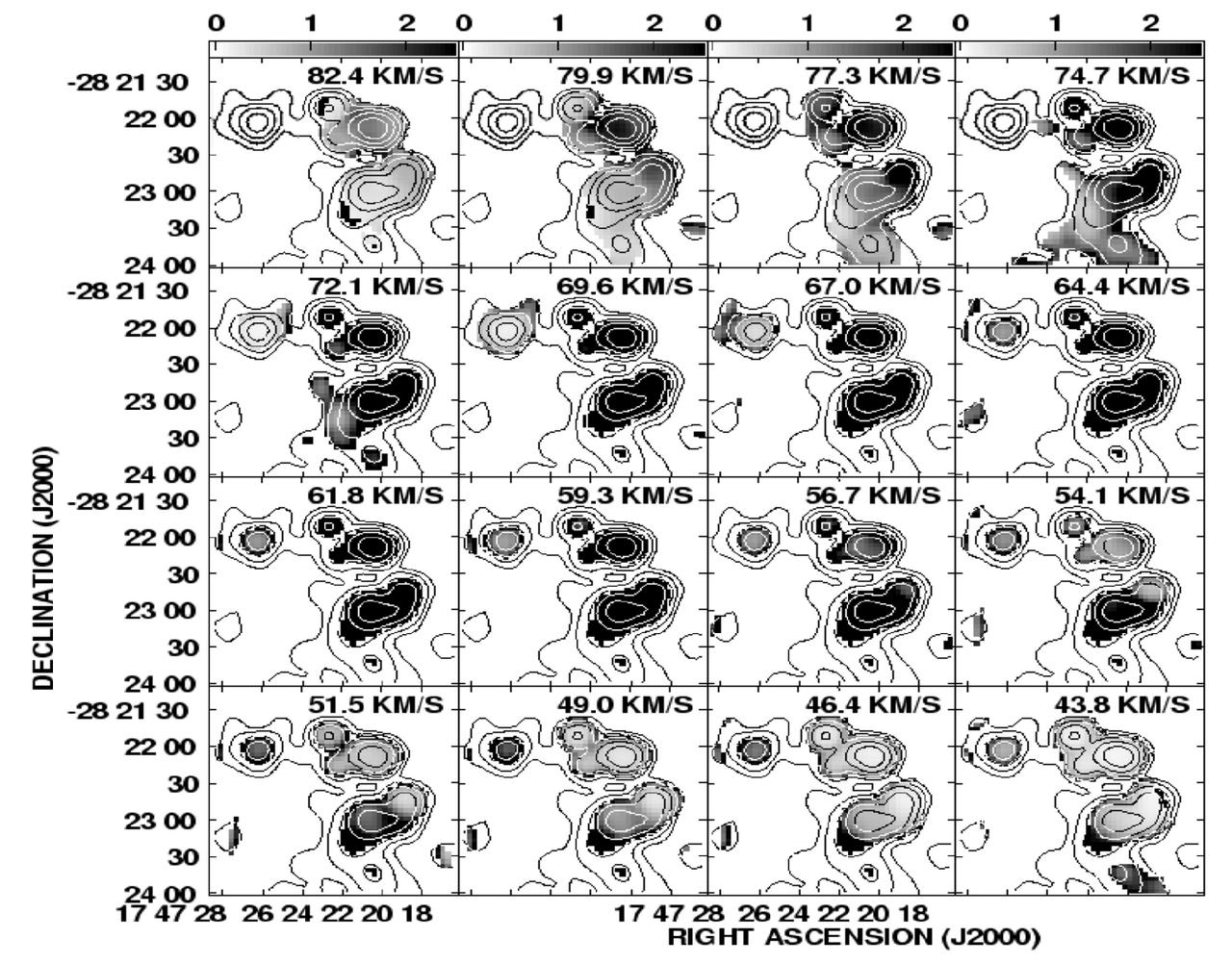}
\caption{Contours representing the continuum toward Sgr B2 (North and Main) and
greyscale represents the HI optical depth for velocities of $\sim$40-80 \kms.
\label{fig18}}
\end{figure}

\begin{figure}
\includegraphics[width=16cm]{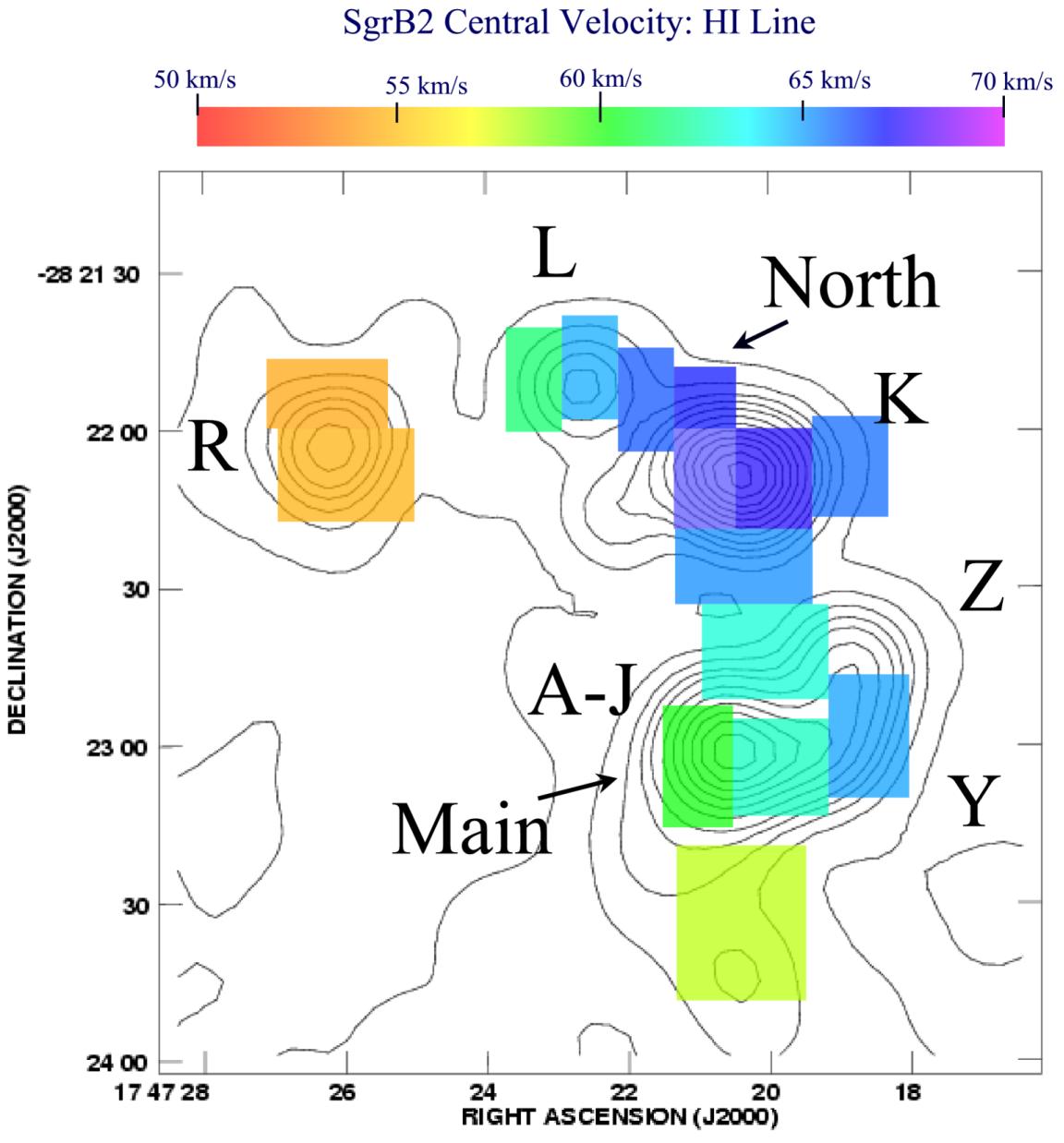}
\caption{Colorscale shows the central velocity of HI absorption across Sgr B2 for small regions where
the profiles have been integrated. The velocities range from $\sim$55 to 68 \kms~and are
overlaid on contours of the 1.4 GHz continuum image shown in Figure 12.
\label{fig19}}
\end{figure}

\begin{figure}
\includegraphics[width=8cm]{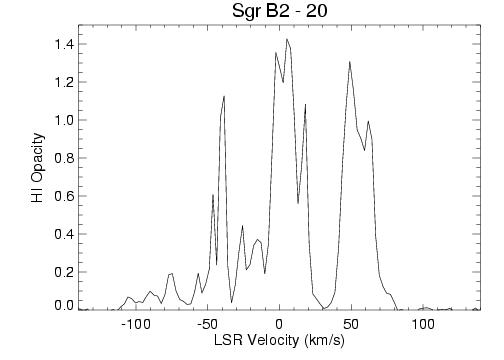} 
\includegraphics[width=8cm]{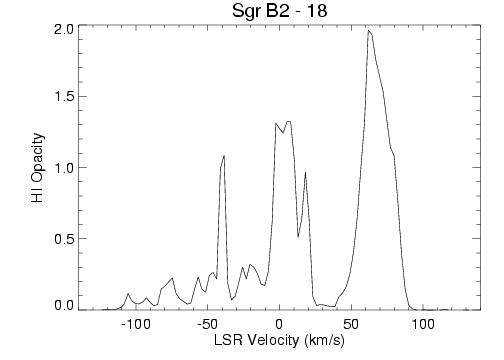} 
\caption{HI opacity as a function of velocity toward (left) Sgr B2-20 and (right) Sgr B2-18
as determined by the VLA. The velocity resolution is 2.5 \kms~and total bandwidth 1.5 MHz.} 
\end{figure}

\begin{figure}
\includegraphics[width=8cm]{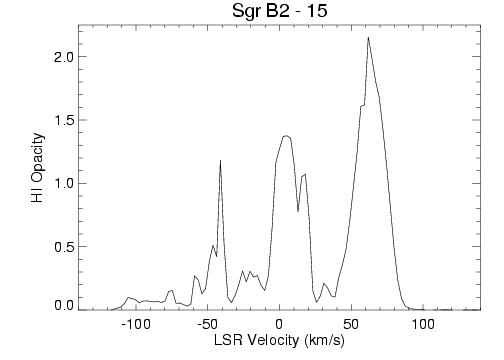} 
\includegraphics[width=8cm]{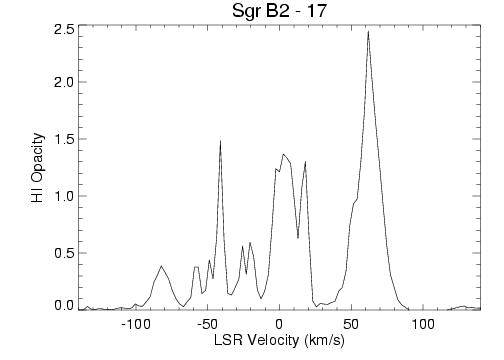} 
\caption{Plot of HI opacity as a function of velocity toward (left) Sgr B2-15 and (right) 
Sgr B2-17 as determined by the VLA. The velocity resolution is 2.5 \kms~and total bandwidth 1.5 MHz.} 
\end{figure}

\begin{figure}
\includegraphics[width=8cm]{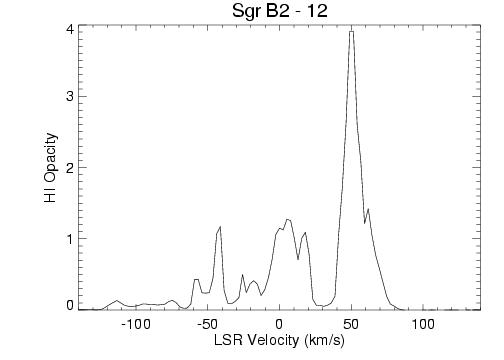} 
\includegraphics[width=8cm]{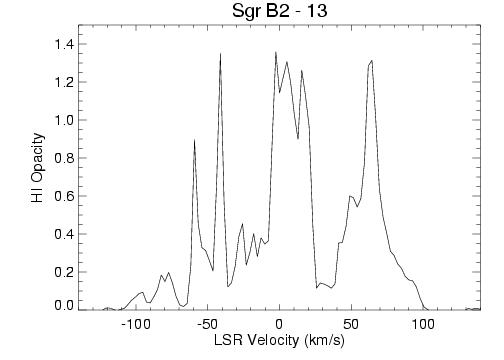} 
\caption{Plot of HI opacity as a function of velocity toward (left) Sgr B2-12 and (right) Sgr B2-18 as determined by the VLA. 
The velocity resolution is 2.5 \kms~and total bandwidth 1.5 MHz.} 
\end{figure}

\begin{figure}
\includegraphics[width=8cm]{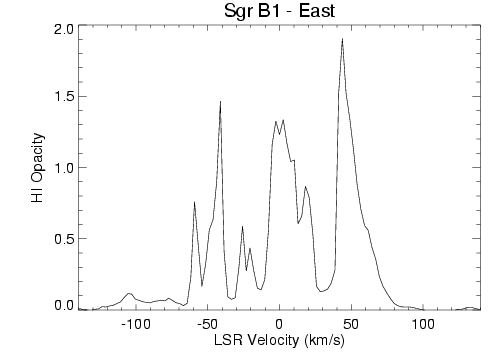} 
\includegraphics[width=8cm]{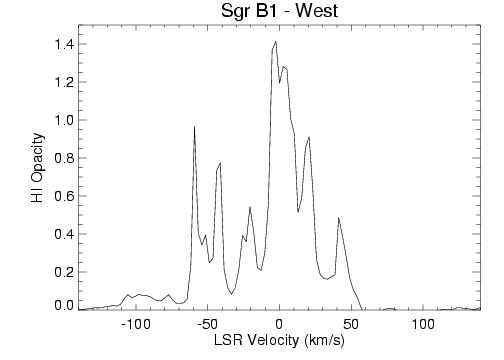} 
\caption{Plot of HI opacity as a function of velocity toward (left) Sgr B1 (East) and (right) Sgr B2 (West) as determined
by the VLA. The small region over which the data were integrated to make these
  profiles is shown in Figure 12. The velocity resolution is 2.5 \kms~and total bandwidth 1.5 MHz.}
\end{figure}

\begin{figure}
\includegraphics[width=16cm]{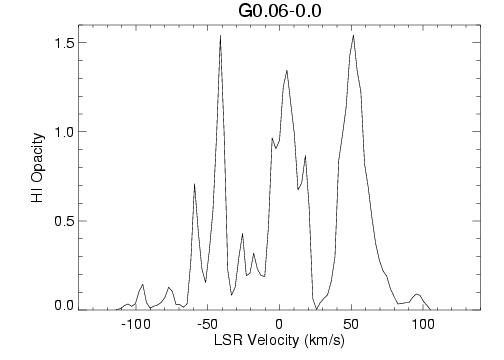}
\caption{Plot of HI opacity toward G0.6-0.0 as determined by the VLA. The velocity resolution is 2.5 \kms~and total bandwidth 1.5 MHz.} 
\end{figure}

\begin{figure}
\includegraphics[width=16cm]{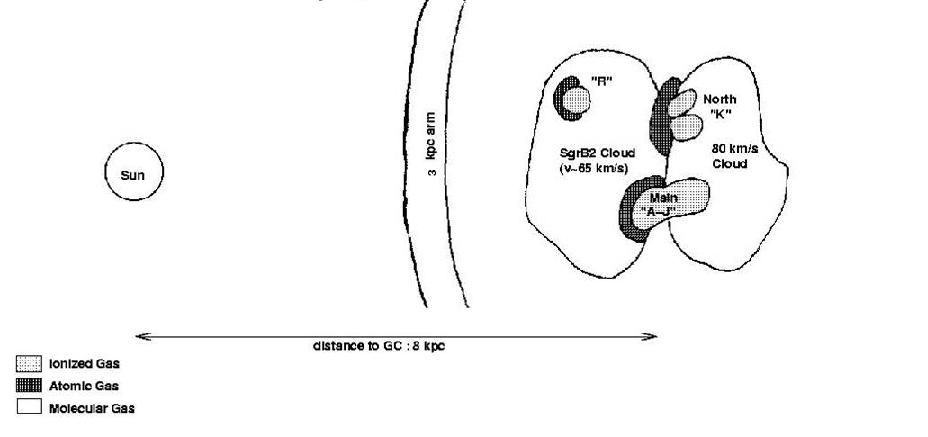}
\caption{Cartoon showing the arrangement of interstellar components in the Sgr B2 complex as derived from comparisons of the spatial and velocity distributions of ionized, molecular and atomic gas. This arrangement illustrates that the compact HII regions in Sgr B2 arise at the interface between two molecular clouds and are likely to be the result of such a collision.} 
\end{figure}

\begin{figure}
\includegraphics[width=16cm]{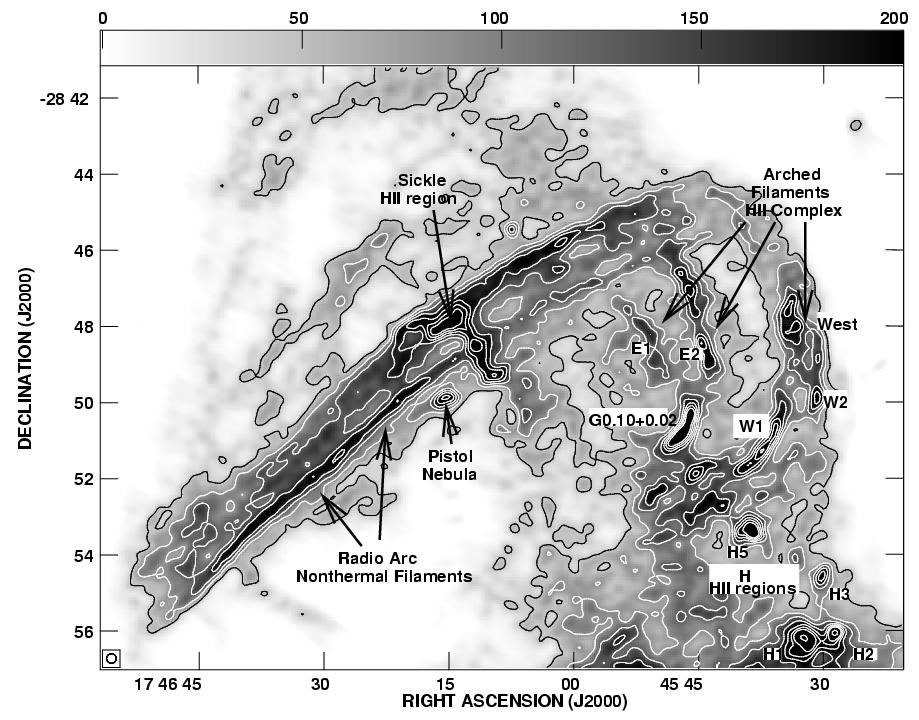}
\caption{VLA 1.4 GHz continuum image of the Radio Arc Region shown in both greyscale and contours. Contour levels represent 5, 10, 15, 20, 25, 30, 35, 50, 100, and 150 times the level of 8.5 mJy \beam. The spatial resolution of the image is 15''.} 
\end{figure}

\begin{figure}
\includegraphics[width=16cm]{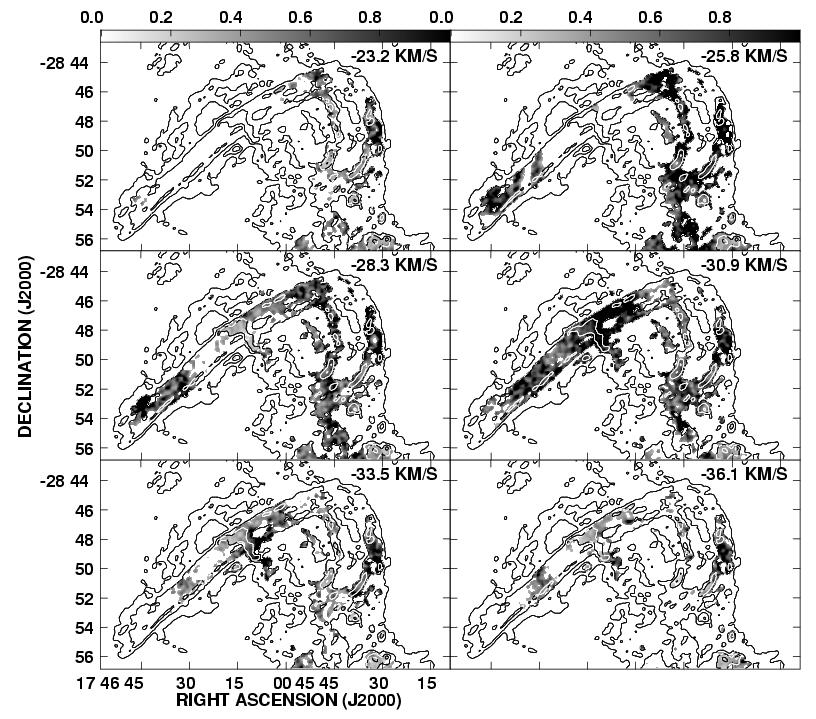} 
\caption{Contours showing 1.4 GHz continuum as in Figure 21, overlaid with greyscale representing the HI opacity at intervals of $\sim$2.5 \kms, beginning with $-$23.2 \kms~and ending with $-$36.1 \kms. The major features in this region (e.g., Arched Filaments HII regions, the H regions, the Sickle, Pistol and Radio Arc non-thermal filaments) are all labeled in Figure 21 for reference.} 
\end{figure}

\clearpage

\begin{figure}
\includegraphics[angle=270,width=16cm]{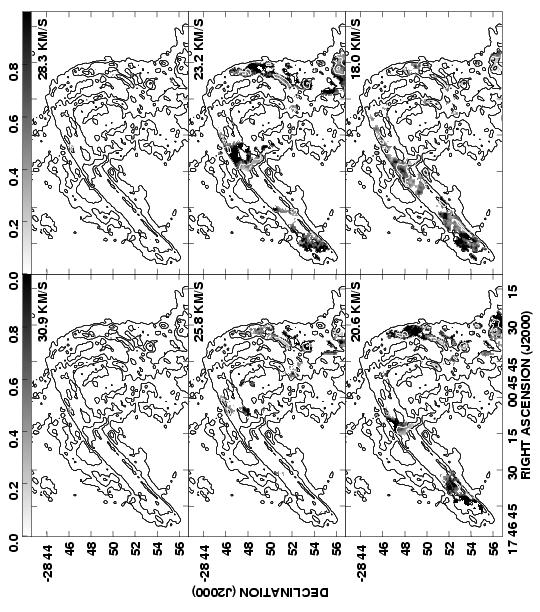} 
\caption{Contours showing 1.4 GHz continuum as in Figure 21, overlaid with greyscale representing the HI opacity at intervals of $\sim$2.5 \kms, beginning with +18.0 \kms~and ending with +30.9 \kms. The major features in this region (e.g., Arched Filaments HII regions, the H regions, the Sickle, Pistol and Radio Arc non-thermal filaments (NTFs)) are all labeled in Figure 21 for reference.} 
\end{figure}

\clearpage

\begin{figure}
\includegraphics[width=8cm]{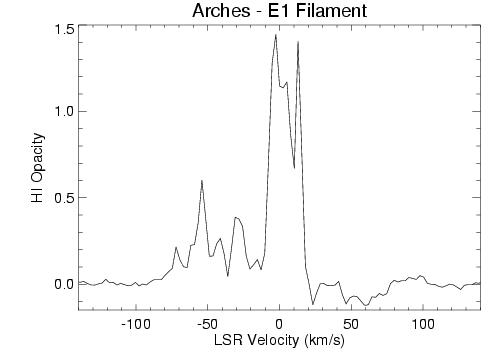} 
\includegraphics[width=8cm]{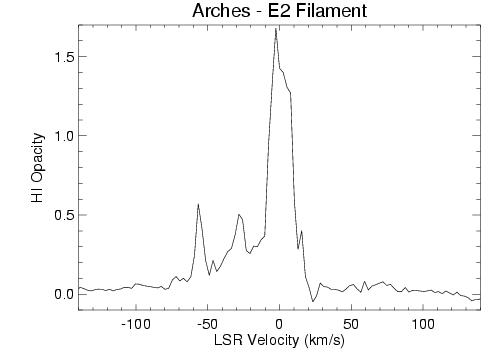} 
\caption{HI opacity towards several regions in the Arched Filaments as determined by the VLA: (left) Arches-E1 and (right) Arches-E2 . Figure 21 shows the locations of these
smaller regions in the Arched Filament complex. The velocity resolution is 2.5 \kms~and total bandwidth 1.5 MHz.}
\end{figure}

\begin{figure}
\includegraphics[width=8cm]{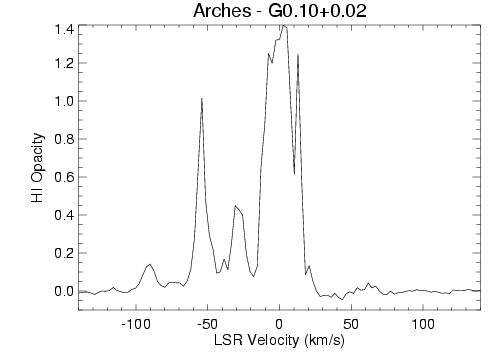} 
\includegraphics[width=8cm]{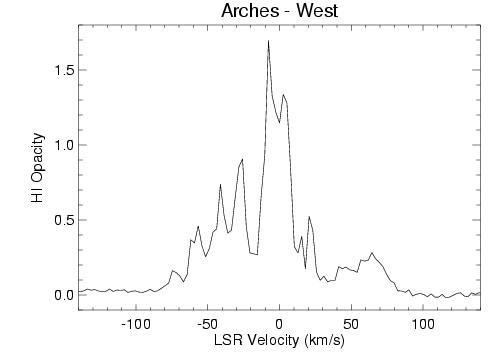} 
\caption{HI opacity towards several regions in Arched Filaments as determined by the VLA: (left) G0.10+0.02 and (right) Arches-West. 
Figure 21 shows the locations of these smaller regions in the Arched Filament complex. The velocity resolution is 2.5 \kms~and bandwidth 1.5 MHz.}
\end{figure}

\begin{figure}
\includegraphics[width=8cm]{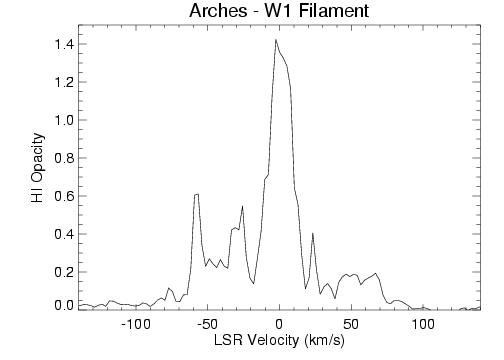} 
\includegraphics[width=8cm]{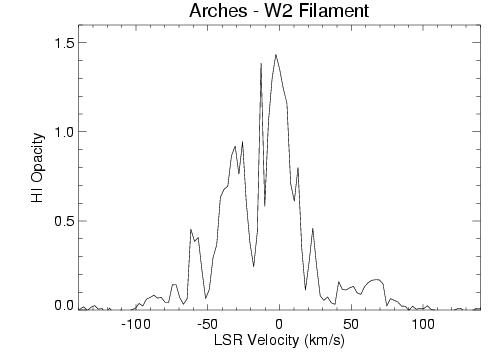} 
\caption{HI opacity towards the Arched Filament (left) Region-W1 and (right) Region-W2, as determined by the VLA. The velocity resolution is 2.5 \kms~and total bandwidth 1.5 MHz.}
\end{figure}

\begin{figure}
\includegraphics[width=16cm]{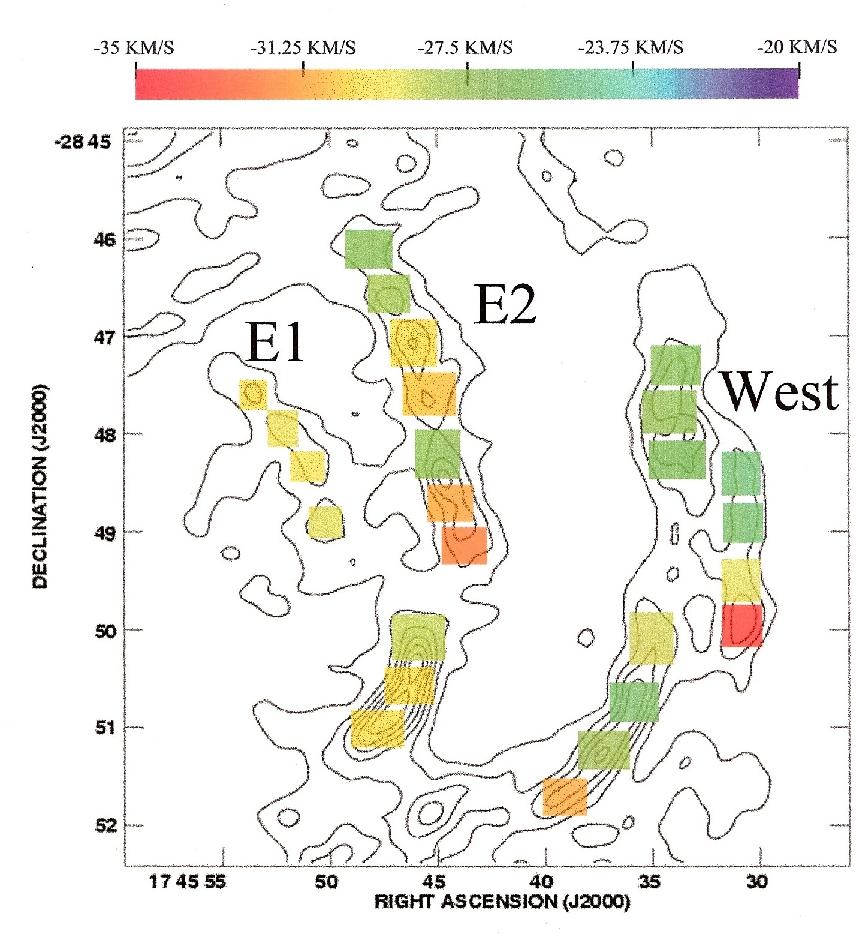} 
\caption{Colorscale shows the central velocity of HI opacity for small regions in the Arched Filaments HII Complex over
which the opacity has been integrated. The velocities range from $\sim$$-$25 to $-$35 \kms~and are
overlaid on contours of the 1.4 GHz continuum image shown in Figure 21.}
\end{figure}

\begin{figure}
\includegraphics[width=16cm]{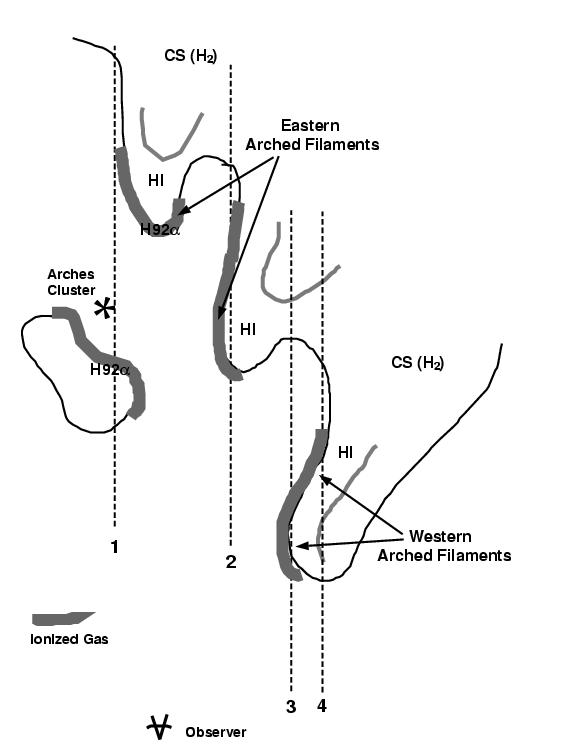} 
\caption{Schematic illustration of the arrangement of the ionized, molecular and atomic gas in the Arched Filaments. Originally published in Lang et al. (2002).} 
\end{figure}

\begin{figure}
\includegraphics[width=8cm]{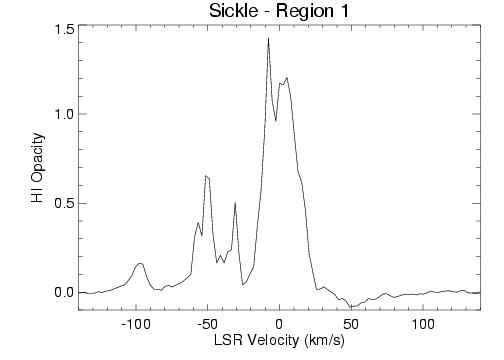} 
\includegraphics[width=8cm]{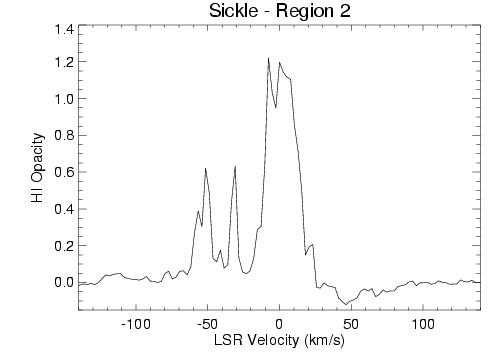} 
\caption{HI opacity towards the Sickle (left) Region 1 (north; centered at RA, DEC (J2000): 17 46 15, $-$28 48 00), 
and (right) Region 2 (middle; centered at RA, DEC (2000): 17 46 10, $-$28 49 00), as determined by the VLA. The velocity resolution is 2.5 \kms~and total bandwidth 1.5 MHz.}
\end{figure}

\begin{figure}
\includegraphics[width=8cm]{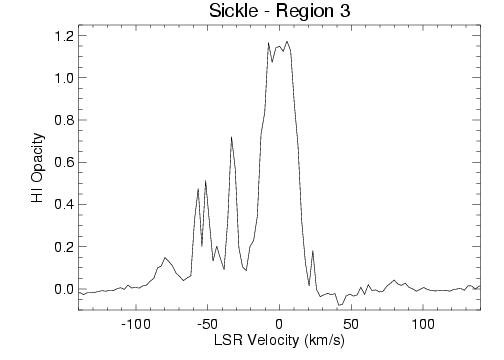} 
\includegraphics[width=8cm]{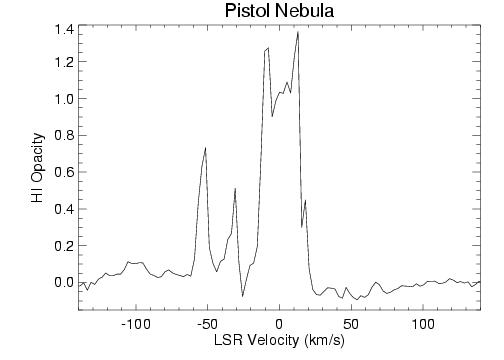} 
\caption{HI opacity towards the Sickle (left) Region 3 (South, centered at RA, DEC (J2000): 17 46 07, $-$28 50 00) and
(right) the Pistol Nebula, as determined by the VLA. The velocity resolution is 2.5 \kms~and total bandwidth 1.5 MHz.}
\end{figure}

\begin{figure}
\includegraphics[width=16cm]{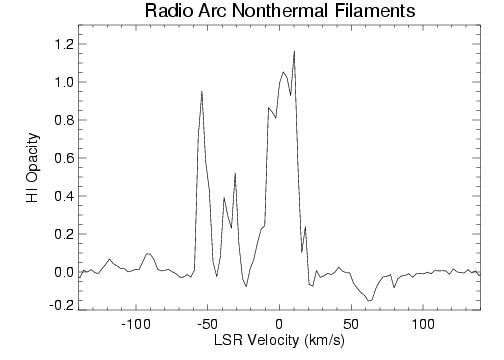} 
\caption{HI opacity toward the Radio Arc non-thermal filaments (NTFs), as determined by the VLA. 
The velocity resolution is 2.5 \kms~and total bandwidth 1.5 MHz.}
\end{figure}

\begin{figure}
\includegraphics[width=8cm]{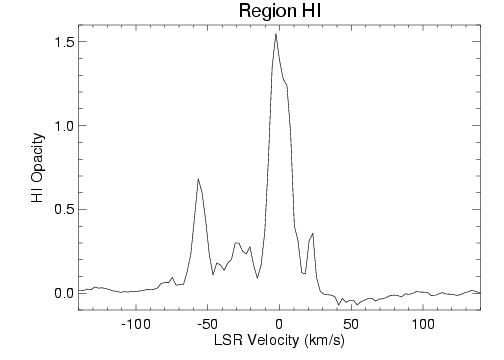} 
\includegraphics[width=8cm]{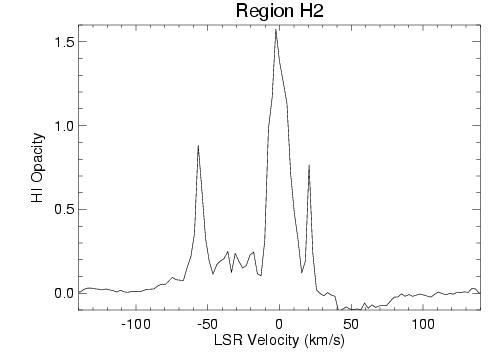} 
\caption{HI opacity towards the HII regions (left) H1 and (right) H2 as determined by the VLA. The velocity resolution is 2.5 \kms~and total bandwidth 1.5 MHz.}
\end{figure}

\begin{figure}
\includegraphics[width=8cm]{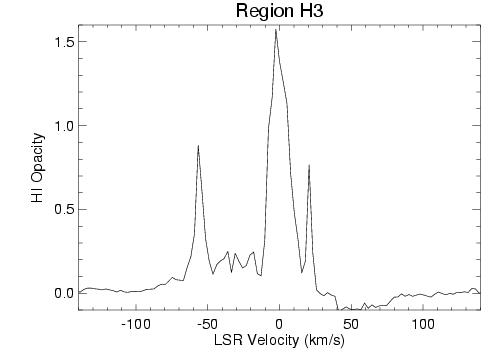} 
\includegraphics[width=8cm]{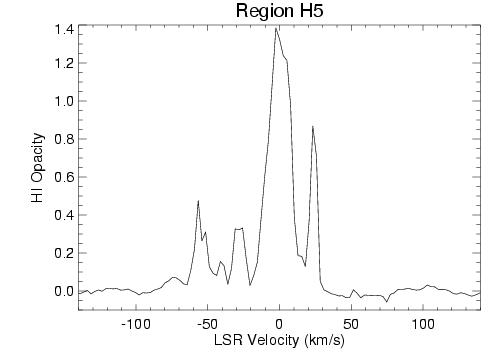}
\caption{HI opacity towards the HII regions (left) H3 and (right) H5 as determined by the VLA. The velocity resolution is 2.5 \kms~and total bandwidth 1.5 MHz.}
\end{figure}

\begin{figure}
\includegraphics[width=16cm]{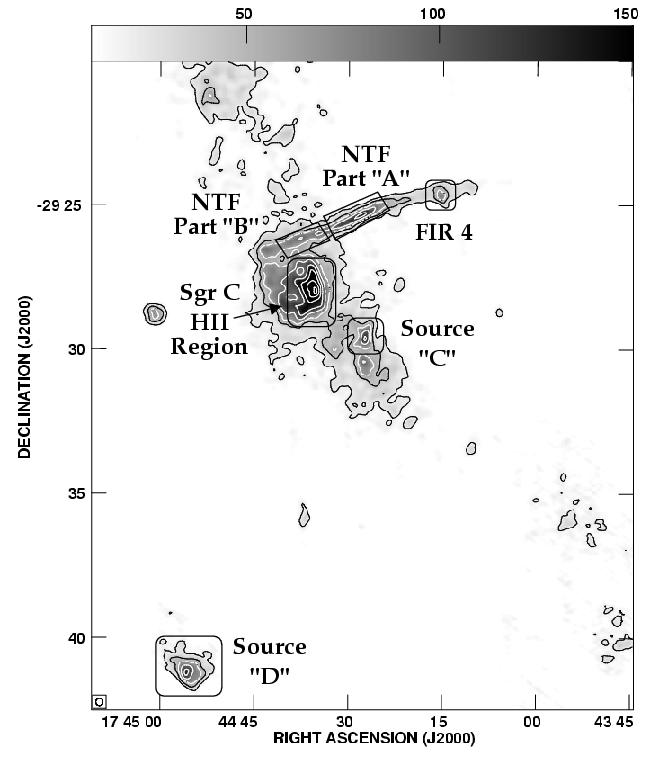} 
\caption{VLA 1.4 GHz continuum image of the region surrounding Sgr C shown in both greyscale and contours. Contour levels represent 10, 20, 30, 45, 60, 75, 90, and 100 percent of the peak flux of 208 mJy/beam. The spatial resolution of the image is 15\arcsec.}
\end{figure}

\begin{figure}
\includegraphics[width=8cm]{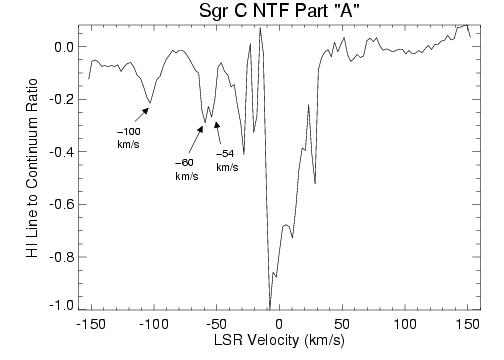} 
\includegraphics[width=8cm]{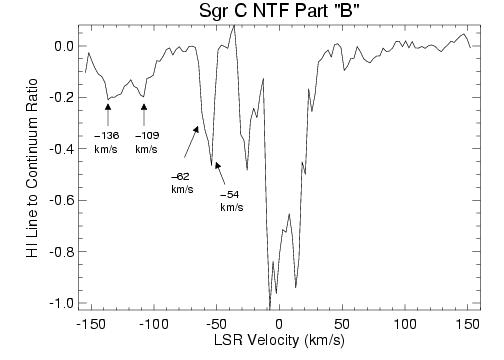}
\caption{HI absorption spectra towards (left) Part A and (right) Part B of the Sgr C NTF as determined by the VLA. The velocity resolution is 2.5 \kms~and total bandwidth 1.5 MHz.}
\end{figure}

\begin{figure}
\includegraphics[width=8cm]{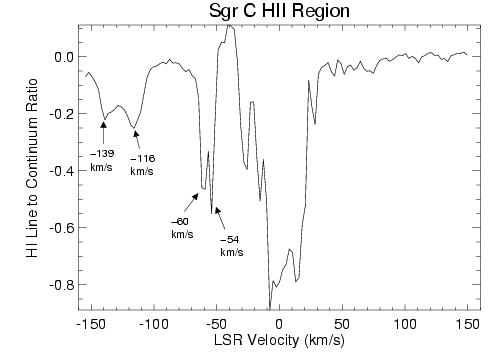} 
\includegraphics[width=8cm]{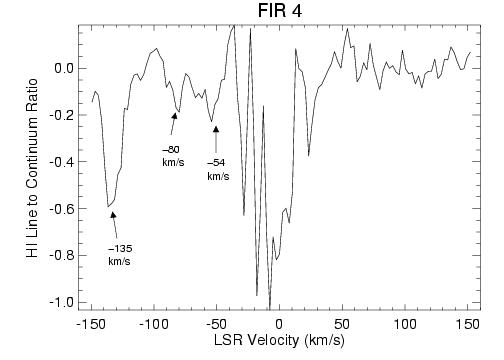}
\caption{HI absorption spectra towards the (left) Sgr C HII Region and (right) FIR 4 as determined by the VLA. The velocity resolution is 2.5 \kms~and total bandwidth 1.5 MHz.}
\end{figure}

\begin{figure}
\includegraphics[width=8cm]{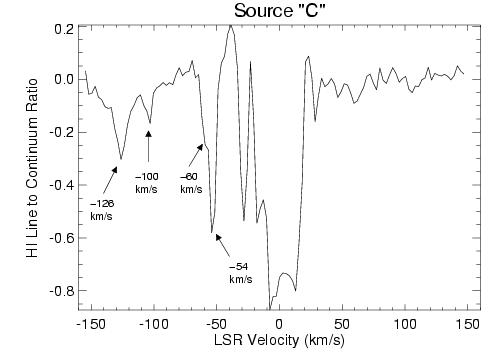} 
\includegraphics[width=8cm]{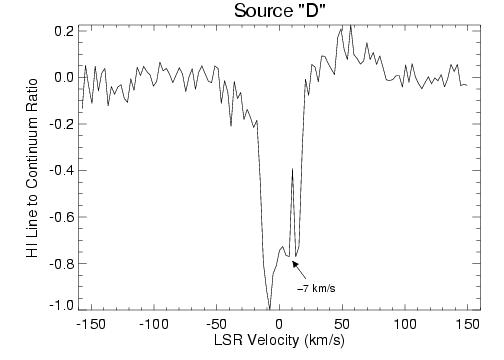}
\caption{HI absorption spectra towards (left) Source C  and (right) Source D as determined by the VLA. The velocity resolution is 2.5 \kms~and total bandwidth 1.5 MHz.}
\end{figure}

\begin{figure}
\includegraphics[width=8cm]{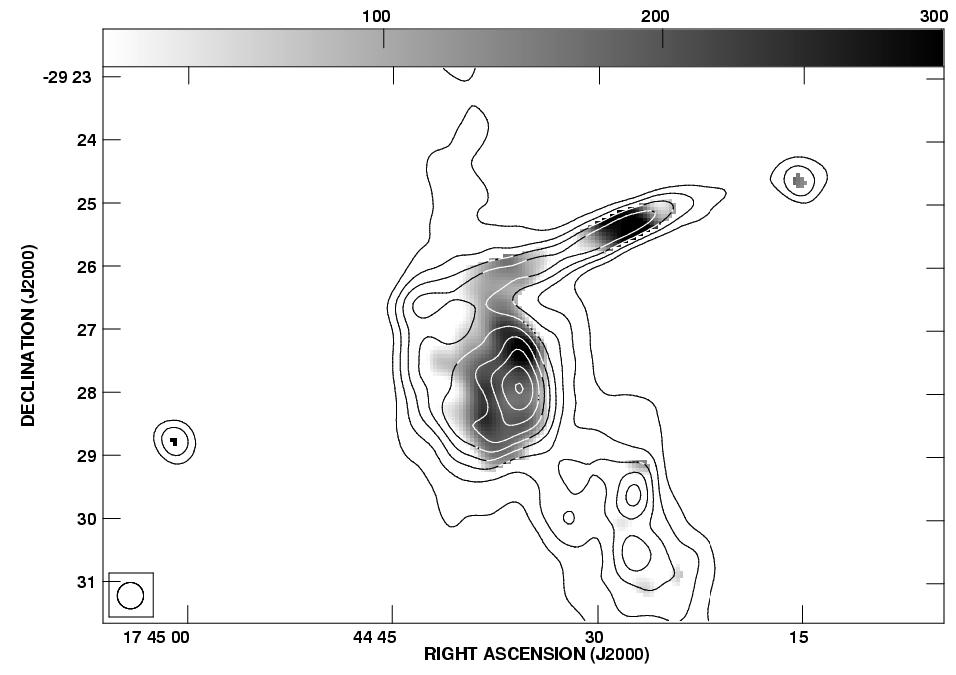} 
\includegraphics[width=8cm]{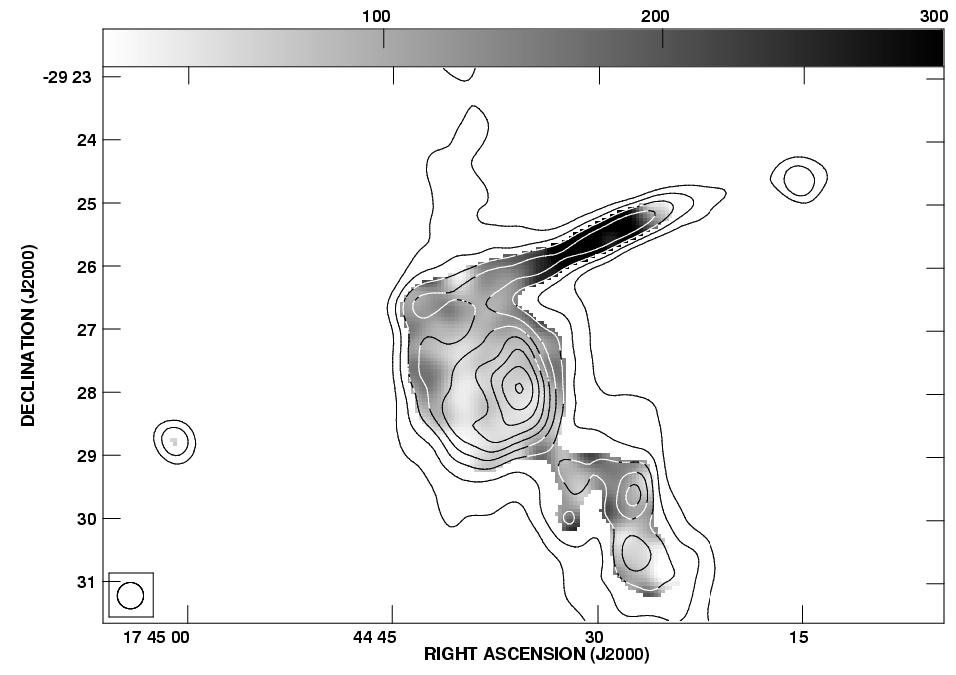} 
\caption{VLA 1.4 GHz continuum contours overlaid on optical depth greyscale (with a range of 0 to 0.3) 
for velocities of (left) $\sim$ $-$65~\kms~and (right) $\sim$$-$100~\kms.}
\end{figure}

\clearpage

\begin{figure}
\includegraphics[width=8cm]{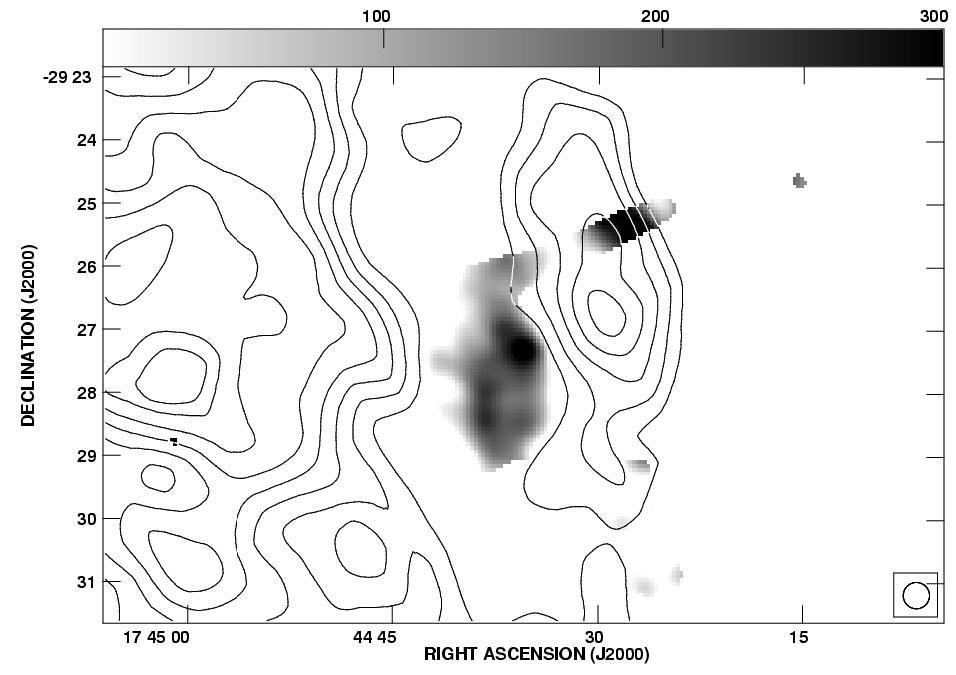} 
\includegraphics[width=8cm]{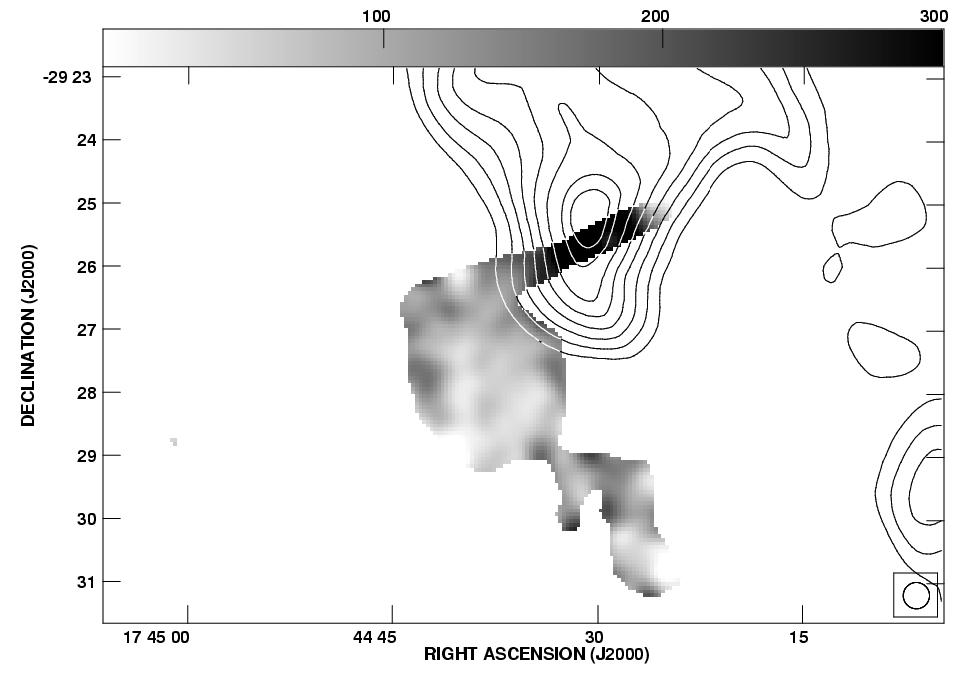} 
\caption{Contours of CO~(J=1-0) emission from Oka et al. (1998) for (left) $\sim$ $-$65~\kms~and 
(right) $\sim$$-$100~\kms~overlaid on the same optical depth greyscale as in Figure 36.}
\end{figure}
\clearpage

\begin{figure}
\includegraphics[width=16cm]{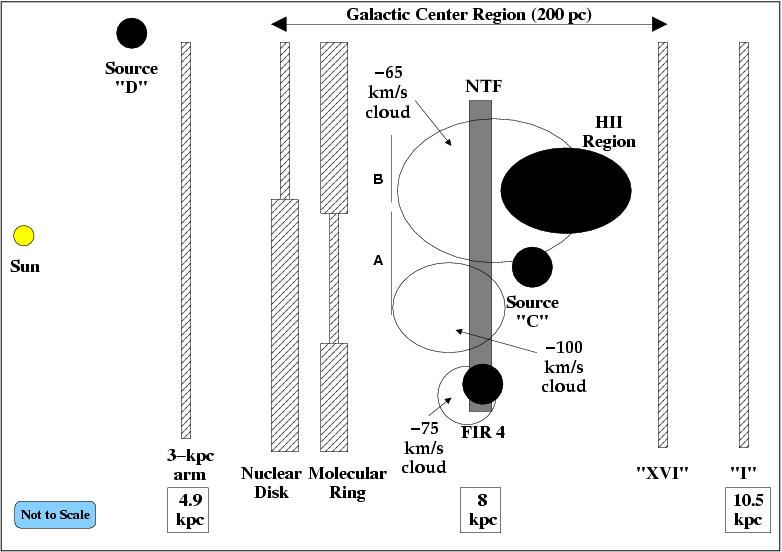} 
\caption{A schematic of the possible arrangement of the Sgr C complex sources and atomic and molecular components.}
\end{figure}

\clearpage

\begin{figure}
\includegraphics[width=16cm]{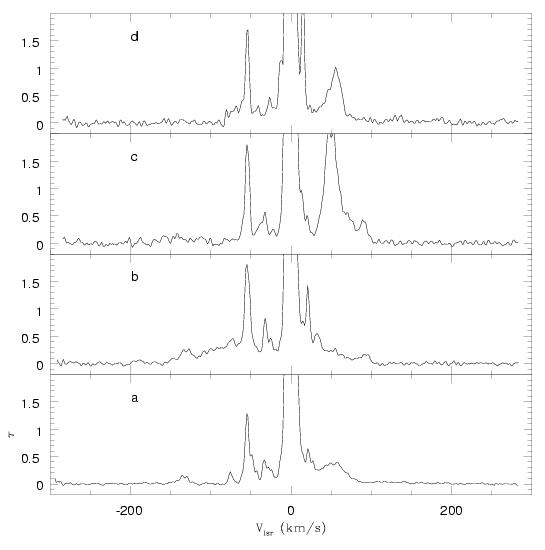} 
\caption{Optical depth spectra from the A+B+C+D array data taken from  Dwarakanath et al. (2004). 
The four spectra correspond to the positions a (SgrA*), b ($\sim$35\arcsec~to the SW of SgrA*), 
c (2\arcmin~to the E of SgrA*; in SgrA East) and d (2\arcmin~to the NE of SgrA*). 
A wide line (FWHM $\sim$ 120 km s$^{-1}$) is clearly detected at position b and is 
evident as a broad shoulder underneath the narrow lines ($\sim$10 \kms) and can 
be associated with the circumnuclear disk.}

\end{figure}

\end{document}